\documentclass[aps,pre,nofootinbib,10pt,superscriptaddress,groupaddress]{revtex4-1}
\usepackage{amsmath,amssymb,tabularx,graphicx}

\DeclareMathOperator\erf{erf}
\usepackage{color}
\newcolumntype{x}{>{\centering\arraybackslash}X}
\newcommand{\tr}[1]{\textcolor{black}{#1}}
\newcommand{\newtr}[1]{\textcolor{black}{#1}}
\begin{document}
\title{EH-DPD: a Dissipative Particle Dynamics approach to Electro-Hydrodynamics}
\author{Alberto Gubbiotti}
\affiliation{Dipartimento di Ingegneria Meccanica e Aerospaziale, Sapienza Universit\`a di Roma, Via Eudossiana 18, 00184 Roma, Italia.}
\email{alberto.gubbiotti@uniroma1.it}
\author{Mauro Chinappi}
\affiliation{Dipartimento di Ingegneria Industriale, Universit\`a di Roma Tor Vergata, Via del Politecnico 1, 00133 Roma, Italia.}
\author{Carlo Massimo Casciola}
\affiliation{Dipartimento di Ingegneria Meccanica e Aerospaziale, Sapienza Universit\`a di Roma, Via Eudossiana 18, 00184 Roma, Italia.}
%%

%%z
\begin{abstract}
Electrohydrodynamics is crucial in many nanofluidic and biotechnological applications. 
In such small scales, the complexity due to the coupling of fluid dynamics with the dynamics of ions is increased by the relevance of thermal fluctuations.
Here, we present a mesoscale method based on the Dissipative Particle Dynamics (DPD) model of the fluid.
Two scalar quantities, corresponding to the number of positive and negative 
ions carried by each DPD particle, are added to the standard DPD formulation.
We introduced a general framework that, given the definition of the free-energy of the DPD particle, allows to derive a fluctuation-dissipation relation and 
the expression for ionic fluxes between the DPD particles. 
This provides a link between the dynamics of the system and its equilibrium properties.
The model is then validated simulating a planar electroosmotic flow for the cases of
overlapping and non overlapping electric double layers.
\newtr{It is shown that using a Van der Waals equation of state the effect of ionic finite size can be accounted, leading to significant 
effects on the concentration and velocity profiles 
with respect to the ideal solution case.}
\end{abstract}
\maketitle
\section{introduction}

Electrohydrodynamics deals with the coupled motion of fluids and ions~\cite{theoretical_microfluidics}.
Such coupling has relevant effects in microfluidics, allowing manipulation of fluids or dispersed particles using electrical stimuli~\cite{electrohydrodynamics_microsystems}.
A specific range of applications in which electrohydrodynamic effects are of crucial importance are nanopore systems~\cite{dekker2007solid,gubbiotti2021electroosmosis}.
When a fluid containing ions is located in a confined region as a nanopore, a non uniform distribution of ions may arise in a region whose size depends on the ionic concentration in the bulk, the so-called Debye layer~\cite{theoretical_microfluidics}.
This inhomogeneity in the ionic distribution may be generated as a consequence of the interaction of the ions with the nanopore walls, as in the case of 
a charged nanopore surface~\cite{ma2019nanopore,laohakunakorn2014electroosmotic}.
However, it may also be induced by ion specific interactions with neutral walls, as has been studied in the case of hydrophobic nanopores~\cite{kim2009high}.
Ionic inhomogeneity can be also achieved via an external gating voltage 
applied to electrodes embedded in the nanopore~\cite{bai2014fabrication,cantley2019voltage}
or via induced charge mechanism where the same external electric field 
that drives the ions through the pore, also polarizes the solid membrane inducing a surface 
potential that, in turn, alters ion distribution in the nanopore~\cite{di2021geometrically,yao2020induced}.
In all these cases, the ionic distributions near the walls are generally different for positive and negative ions.
As a consequence, the zone near the confining walls is electrically charged, and the fluid in that region can be put in motion by an external electric field.
The resulting flow is known as electroosmotic flow and it has been shown to take place both in synthetic~\cite{yusko2009electroosmotic,laohakunakorn2014electroosmotic,balme2015ionic} and in biological~\cite{bonome2017electroosmotic,boukhet2016probing,huang2017electro} nanopores.
Electroosmotic flow can be the dominant effect governing the translocation of particles or molecules through a nanopore~\cite{malgaretti2014entropic,asandei2016electroosmotic,boukhet2016probing,huang2020electro}, and can generate interesting phenomena such as current rectification~\cite{yusko2009electroosmotic} or complex velocity profiles~\cite{chinappi2018charge,chinappi2020analytical}.

In all the mentioned examples, the modeling of confined systems has to combine electrohydrodynamic phenomena with thermal fluctuations, which are especially important in nanopores~\cite{marbach2021intrinsic}.
For this reason, an extensively used technique to simulate nanopores is all-atoms Molecular Dynamics (MD)~\cite{maffeo2012modeling,bonome2017electroosmotic,chinappi2018charge} that naturally includes all the relevant effects.
For systems out of the typical length and time scales accessible to Molecular Dynamics, mesoscale models which reduce the degrees of freedom while properly modeling the thermal fluctuations of the system are needed, for a review on computational methods to study electrohydrodynamics at the nanoscale, see,
among others~\cite{rotenberg2013electrokinetics,gubbiotti2021electroosmosis}.

A technique which has been widely used to simulate mesoscale systems is Dissipative Particle Dynamics (DPD)~\cite{espanol2017perspective}.
In the DPD framework, the fluid is represented by a system of pairwise interacting particles~\cite{hoogerbrugge1992simulating,espanol1995statistical}.
The original DPD model was developed to study the rheology of colloidal suspensions~\cite{hoogerbrugge1992simulating,bolintineanu2014particle}, but in the last two decades it has been expanded in many different ways in order to simulate increasingly complex systems~\cite{espanol2003smoothed,pagonabarraga2001dissipative,avalos1997dissipative,espanol1997dissipative,li2015transport,deng2016cdpd}. 
Physical systems studied with DPD or derived methods include blood~\cite{katanov2015microvascular,blumers2017gpu}, polymers~\cite{kreer2016polymer}, biomolecules~\cite{peter2015polarizable}, biological membranes~\cite{sevink2014efficient}, and droplets~\cite{wang2015droplets}.
DPD simulations including electrostatic interactions have also been performed, 
either considering DPD particles with fixed charge~\cite{groot2003electrostatic,gonzalez2006electrostatic,smiatek2011mesoscopic} or charged polyelectrolytes~\cite{sindelka2014dissipative,lisal2016self}.
A different approach considers the concentrations of ionic species as additional scalar variables associated to each DPD particle, modelling the fluxes of concentration between them~\cite{deng2016cdpd}. \tr{This approach is promising since it allows to rescale the system size without having to explicitly parametrize the ion-solvent interaction. However, a link between the equation of state of the electrolyte solution and the ionic transport equations is still lacking, restricting the fluxes which can be simulated to advection plus Fickian diffusion.}

Here, we propose a mesoscale model based on DPD which is able to simulate the electrohydrodynamic phenomena taking place in nanofluidic systems. We will refer to this method as electrohydrodynamic-DPD, or EH-DPD. 
\footnote{We refer to electrohydrodynamics in its broader meaning of phenomena involving charge transport coupled to fluid motion~\cite{saville1997electrohydrodynamics}.}
\newtr{The dissolved ions are represented by adding two degrees of freedom for each meso-particle, and the exchange of ions depends on the difference of chemical potential}.
Although only two charged species are considered here, the model can be easily generalized to include electrolyte solutions with more species.
In section~\ref{sec:equations} the equations for the dynamics of the system are reported, and it is shown (section~\ref{sec:equilibrium}) that, if appropriate fluctuation-dissipation conditions are satisfied, the proposed dynamics admits an equilibrium distribution.
Since the equilibrium distribution of a system is related to its thermodynamic potential, this gives a link between the terms arising in the equations of motion and the thermodynamic properties of the meso-particles, allowing a consistent definition of pressure and chemical potential.
In section~\ref{sec:model} the physical model used to derive the forces and the ionic exchange rates between meso-particles is described.
The electrostatic interactions are computed considering the charge carried by each meso-particle to be distributed as a Gaussian of constant variance located on its center.
The chemical potential used in the model is that of a perfect gas plus a contribution due to the electrostatic interactions.
The ionic exchange rates between the meso-particles are modelled as dependent on the local ionic quantities to obtain a conductance which is linearly dependent on the ionic average concentration.
It is shown that this dependence implies the necessity of considering an additional drift in order for the system to reach the desired equilibrium distribution.
In section~\ref{sec:validation} the model is tested against analytical results for planar electroosmotic flow, finding an excellent agreement with the theoretical prediction for
both the cases of overlapping and non overlapping electric double layers.
\newtr{As an example of the applicability of the presented approach to simulate a more complex fluid, a Van der Waals equation of state is also used, simulating ion-specific effects such as excluded volume.}
The possibility of simulating different equations of state for the electrolyte solution is promising for the study of current and mass transport in systems in which phase transitions and ion specific effects are relevant, such as hydrophobic nanopores and hydrophobic nanporous materials~\cite{tinti2017intrusion,camisasca2020gas,trick2017voltage,polster2020gating}.

\section{Electro-Hydrodynamics: DPD formulation}
\label{sec:equations}
The fluid is constituted by $N$ meso-particles of equal mass $m$.
The state of the $i^{\mathrm{th}}$ meso-particle is described by its position $\boldsymbol{x}_i$, velocity $\boldsymbol{v}_i$, quantity of cations $n^c_i$ and quantity of anions $n^a_i$.
The vector of state of the entire systems has therefore dimension $8N$ and the equations for its evolution are
%
% EHDPD_1
\begin{align}
% eq1
\label{eq:ehdpd_1}
&\mathrm{d}\boldsymbol{x}_i=
\boldsymbol{v}_i\mathrm{d}t\;,\\
% eq2
\label{eq:ehdpd_2}
&m\mathrm{d}\boldsymbol{v}_i=\boldsymbol{f}^{\boldsymbol{C}}_{i}\mathrm{d}t+
\sum\limits_{j\ne i}\left[
\gamma w^D_{ij}\boldsymbol{v}_{ji}\cdot\boldsymbol{e}_{ij}\mathrm{d}t+
\sigma w^R_{ij}\mathrm{d}W^v_{ij}
\right]\boldsymbol{e}_{ij}\;,\\
% eq3
\label{eq:ehdpd_3}
&\mathrm{d}n^c_i=
\sum\limits_{j\ne i}\left[
\gamma^c w^D_{ij}h^{c}_{ij}\mathrm{d}t+
\sigma^c w^R_{ij}\mathrm{d}W^c_{ij}\right]\;,\\
% eq4
\label{eq:ehdpd_4}
&\mathrm{d}n^a_i=
\sum\limits_{j\ne i}\left[
\gamma^aw^D_{ij} h^{a}_{ij}\mathrm{d}t+
\sigma^aw^R_{ij}\mathrm{d}W^a_{ij}\right]\;,
\end{align}
where $\boldsymbol{e}_{ij}=\left(\boldsymbol{x}_i-\boldsymbol{x}_j\right)/|\boldsymbol{x}_i-\boldsymbol{x}_j|$ is the unit vector pointing along the particle-particle direction.
The increments $\mathrm{d}W_{ij}$ are independent increments of the Wiener process, three for each pair of particles, satisfying 
\begin{align}
\label{eq:dWv}
&\mathrm{d}W^{v}_{ij}=\mathrm{d}W^{v}_{ji}\;,\\
\label{eq:dWc}
&\mathrm{d}W^{c}_{ij}=-\mathrm{d}W^{c}_{ji}\;,\\
\label{eq:dWa}
&\mathrm{d}W^{a}_{ij}=-\mathrm{d}W^{a}_{ji}\;.
\end{align}

Equations~\eqref{eq:ehdpd_1} and \eqref{eq:ehdpd_2} have the structure of the standard DPD equations~\cite{groot1997dissipative}, where $\boldsymbol{f}^{\boldsymbol{C}}_{i}$ is a conservative force which depends on the physical model chosen and will be specified in Section~\ref{sec:model}. 
The parameters $\gamma$ and $\sigma$ control the intensity of the respective forces, i.e. the dissipative force
\begin{equation}
\boldsymbol{f}^{\boldsymbol{D}}_{ij}=\gamma w^D_{ij}\left(\boldsymbol{v}_{ji}\cdot\boldsymbol{e}_{ij}\right)\boldsymbol{e}_{ij}\;,
\end{equation}
 and the random force
\begin{equation}
\boldsymbol{f}^{\boldsymbol{R}}_{ij}=\sigma w^R_{ij}\xi^{v}_{ij}\boldsymbol{e}_{ij}\;,
\end{equation}
where $\boldsymbol{v}_{ji}=\boldsymbol{v}_j-\boldsymbol{v}_i$ is the velocity difference between two interacting meso-particles and $\xi^{v}_{ij}$ is a white noise stochastic process such that $\mathrm{d}W^{v}_{ij}=\xi^{v}_{ij}\mathrm{d}t$.
The functions $w^D$ and $w^R$ are weight functions which depend only on the interparticle distance $r_{ij}=|\boldsymbol{x}_i-\boldsymbol{x}_j|$.
Such weight functions are maximum for $r_{ij}=0$, and vanish if the interparticle distance is larger than a cutoff radius $r_c$.
There is no prescribed functional form for $w^D$ and $w^R$, here the Lucy function~\cite{espanol2003smoothed} is used for $w^D$,
% WEIGHT
\begin{equation}\label{eq:wd}
w^D_{ij}=w^D(r_{ij})=
	\left(1+3\frac{r_{ij}}{r_c}\right)\left(1-\frac{r_{ij}}{r_c}\right)^3\;,
\end{equation}
for $r_{ij}<r_c$ while the other weight function is such that $w^R_{ij}=\left(w^D_{ij}\right)^{1/2}$.

Equations~\eqref{eq:ehdpd_3} and~\eqref{eq:ehdpd_4} represent the main novelty of the present paper and
describe the rate at which the quantity of cations (and anions) carried by the meso-particle change, $\dot{n}_i^c$ (and $\dot{n}_i^a$) respectively.
This rate is the sum of the contributions from all the pairs, and can be divided in two terms, the dissipative rates
\begin{align}
\label{eq:dissipative_rate}
J^D_{ij}=\gamma^cw^D_{ij}h^c_{ij}\;,
%A^D_{ij}=\gamma^aw^D_{ij}h^a_{ij}\;,
\end{align}
and the random rates
\begin{align}
J^R_{ij}=\sigma^cw^R_{ij}\xi^c_{ij}\;,
%A^R_{ij}=\sigma^aw^R_{ij}\xi^a_{ij}\;.
\end{align}
where analogous expressions hold for the anions.
The same weight functions employed in Eq.~\eqref{eq:ehdpd_2}, $w^D$ and $w^R$ are used.
The quantities $\gamma^c$, $\sigma^c$ ($\gamma^a$, $\sigma^a$) determine the magnitude of the dissipative and random rate at which two meso-particles exchange cations (anions).
The random rates depend on the white noise processes $\xi^c_{ij}$ and $\xi^a_{ij}$, corresponding to the Wiener increments $\mathrm{d}W^c_{ij}=\xi^c_{ij}\mathrm{d}t$ and $\mathrm{d}W^a_{ij}=\xi^a_{ij}\mathrm{d}t$.
The quantities $h^c_{ij}$ and $h^a_{ij}$ determine the dissipative rates for the cations and anions, as a function of the state of the system.
As the conservative force $\boldsymbol{f}^{\boldsymbol{C}}_{i}$, also the quantities $h^c_{ij}$ and $h^a_{ij}$ are specified in Section~\ref{sec:model} and will be shown to be related to the chemical potentials of the meso-particle for the cations and the anions, $\mu^c$ and $\mu^a$ respectively.
The conditions of Eq.~\eqref{eq:dWc} and Eq.~\eqref{eq:dWa} imply that $J^R_{ij}=-J^R_{ji}$ for both ionic species.
Assuming that the additional conditions
\begin{align}\label{eq:h_antisymm}
&h^c_{ij}=-h^c_{ji}\\
&h^a_{ij}=-h^a_{ji}
\end{align}
are satisfied, we have also $J^D_{ij}=-J^D_{ji}$.
If such conditions hold, an important consequence is that the total quantity of both species, and hence the total charge of the system, is strictly conserved during the dynamics.
The dynamics of ionic fluxes between particles is sketched in Fig.~1. %\ref{fig:ion_sketch}.
% ION FLUX SKETCH
\begin{figure}
\label{fig:ion_sketch}
\includegraphics[width=0.5\linewidth]{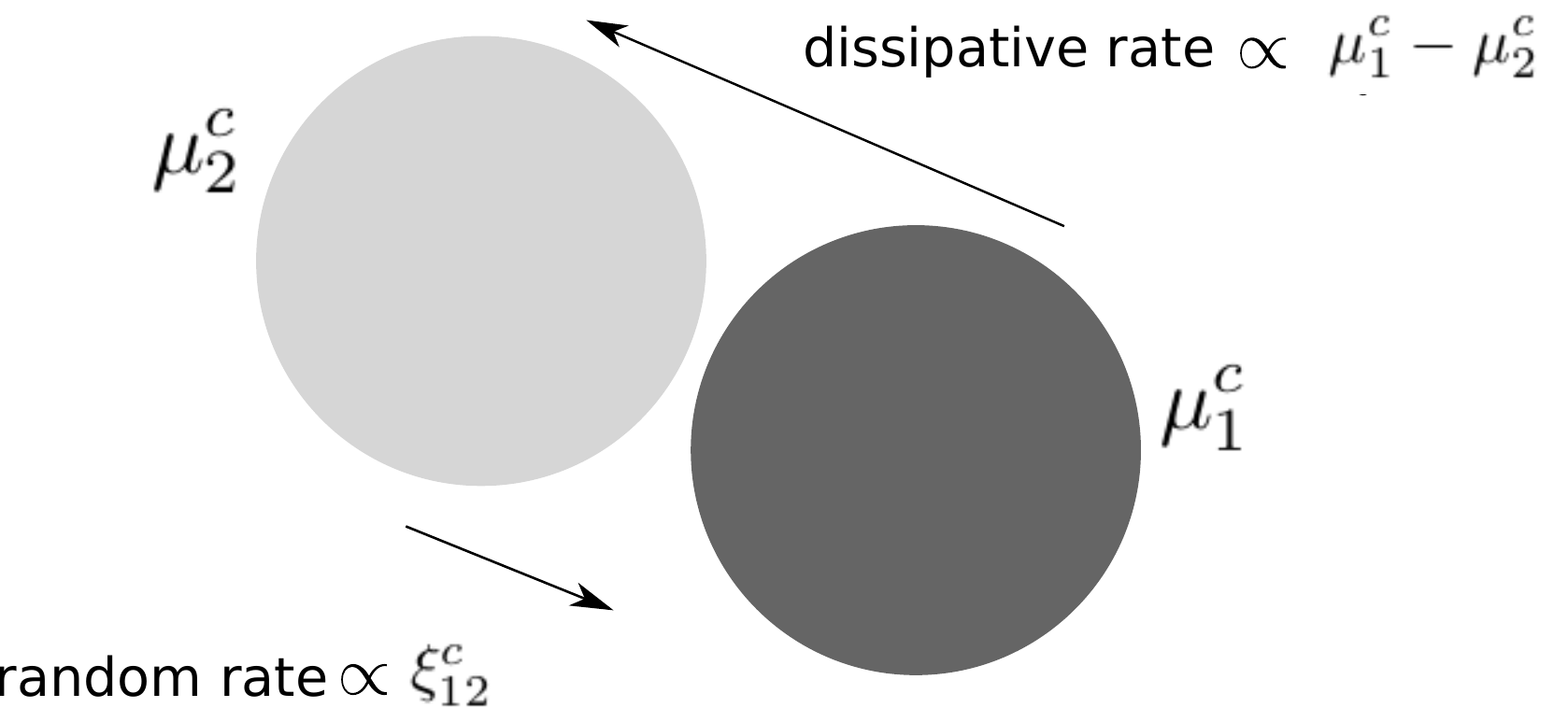}
\caption{
Sketch of the exchange of positive ions $n^c$ between two meso-particles. 
The two particles, labeled $1$ and $2$, have different chemical potentials $\mu^c$, due to different ion concentration and electrostatic potentials. The relation between the chemical potential of the meso-particles and the quantities $h^c_{ij}$ are given in Section~\ref{sec:equilibrium}. 
Two different types of fluxes arise, one dissipative flux proportional to the difference of chemical potential, and one random flux proportional to a white noise process. 
The same applies to negative ions $n^a$.
}
\end{figure}

\section{Equilibrium distribution and Fluctuation Dissipation relations (FDR)}
\label{sec:equilibrium}

The system of Equations~(\ref{eq:ehdpd_1}-\ref{eq:ehdpd_4}) can be written in the compact form of a Langevin equation
% LANGEVIN EQUATION
\begin{equation}
\label{eq:langevin_ehdpd}
\boldsymbol{\dot{y}}=\boldsymbol{u}(\boldsymbol{y})+\boldsymbol{G}(\boldsymbol{y})\boldsymbol{\xi}\;,
\end{equation}
where
%\begin{equation}
%\label{y-def}
%\boldsymbol{y}=\left(\boldsymbol{x},\boldsymbol{v},\boldsymbol{n^c},\boldsymbol{n^a}\right)^T
%\end{equation}
\begin{equation}
\label{y-def}
\boldsymbol{y}=
\begin{pmatrix}
\boldsymbol{x}_1\\ 
\ldots \\
\boldsymbol{x}_N\\
\boldsymbol{v}_1\\ 
\ldots \\
\boldsymbol{v}_N\\
n^c_1\\ 
\ldots \\
n^c_N\\
n^a_1\\ 
\ldots \\
n^a_N\\
\end{pmatrix}
\end{equation}
is the state vector which has dimension $8N$. The standard hydrodynamics setting, with no ion transport, is recovered by reducing the state vector to its 
x- and v-components. In the following, vectors such as $\boldsymbol{u}$ are identified by lowercase bold letters, while matrices as $\boldsymbol{G}$ are identified by uppercase bold letters.

The drift term $\boldsymbol{u}=\left(\boldsymbol{u}^x,\boldsymbol{u}^v,\boldsymbol{u}^c,\boldsymbol{u}^a\right)^T$ includes all the deterministic terms in the equations ~(\ref{eq:ehdpd_1}-\ref{eq:ehdpd_4}), i.e.
% DRIFT DEFINITION
\begin{equation}
\label{u-def}
\boldsymbol{u}(\boldsymbol{y})=\begin{pmatrix}
\boldsymbol{v}_1\\ 
\ldots \\
\boldsymbol{v}_N\\
m^{-1}\left[\boldsymbol{f}^C_{1}+\sum\limits_{j\ne 1}
\gamma w^D_{1j}\left(\boldsymbol{v}_{j1}\cdot\boldsymbol{e}_{1j}\right)\boldsymbol{e}_{1j}\right]\\
\ldots\\
m^{-1}\left[\boldsymbol{f}^C_{N}+\sum\limits_{j\ne N}
\gamma w^D_{Nj}\left(\boldsymbol{v}_{jN}\cdot\boldsymbol{e}_{Nj}\right)\boldsymbol{e}_{Nj}\right]\\
\sum\limits_{j\ne 1}
\gamma^cw^D_{1j}h_{1j}^c\\
\ldots \\
\sum\limits_{j\ne N}
\gamma^cw^D_{Nj}h_{Nj}^c\\
\sum\limits_{j\ne 1}
\gamma^aw^D_{1j}h_{1j}^a \\
\ldots \\
\sum\limits_{j\ne N}
\gamma^aw^D_{Nj}h_{Nj}^a
\end{pmatrix}\;.\end{equation}
The stochastic vector 
%\begin{equation}
%\label{w-def}
%\boldsymbol{\xi}=\left(\boldsymbol{\xi}^v,\boldsymbol{\xi}^c,\boldsymbol{\xi}^a\right)^T
%\end{equation}
\begin{equation}
\label{w-def}
\boldsymbol{\xi}=
\begin{pmatrix}
\xi^v_{12}\\ 
\ldots \\
\xi^v_{(N-1)N}\\
\xi^c_{12}\\ 
\ldots \\
\xi^c_{(N-1)N}\\
\xi^a_{12}\\ 
\ldots \\
\xi^a_{(N-1)N}\\
\end{pmatrix}
\end{equation}
is composed of independent white noise processes and has dimension $3N(N-1)/2$, i.e. three times the total number of particle pairs.
The matrix $\boldsymbol{G}$ has therefore dimension $8N\times 3N(N-1)/2$, and is composed by the following blocks
% NOISE INTENSITY MATRIX BLOCKS
\begin{equation}
\label{g-def}
\boldsymbol{G}(\boldsymbol{y})=\begin{pmatrix}
\boldsymbol{0} & \boldsymbol{0} & \boldsymbol{0} \\
\boldsymbol{G^v} & \boldsymbol{0} & \boldsymbol{0} \\
\boldsymbol{0} & \boldsymbol{G^c} & \boldsymbol{0} \\
\boldsymbol{0} & \boldsymbol{0} & \boldsymbol{G^a}
\end{pmatrix}\;.\end{equation}
The matrix $\boldsymbol{G^v}$ has dimension $3N\times N(N-1)/2$ and can be written in a compact form considering it to be composed of $N\times N(N-1)/2$ vectors of dimension 3, $\boldsymbol{g}^v_{i\alpha}$, each one containing the stochastic force acting on particle $i$ due to the process $\xi^v_\alpha$, where $\alpha$ is an index which spans all the particle pairs, i.e. $\alpha=\alpha(p,q)$ with $p\in[1,N-1]$ and $q\in[p+1,N]$.
Hence, $r_\alpha=r_{pq}$, $w^R_\alpha=w^R_{ij}$, $\boldsymbol{e}_\alpha=\boldsymbol{e}_{pq}$, and, using this compact notation,  $\boldsymbol{g}^v_{i\alpha}=m^{-1}f_{i\alpha}\sigma w^R_\alpha\boldsymbol{e}_\alpha$, where
% F_ia DEFINITION
\begin{equation}
f_{i\alpha}=\begin{cases}
\quad 0 &\quad\mathrm{if}\,i\ne p\,\mathrm{and}\,i\ne q\;,\\
\quad 1 &\quad\mathrm{if}\,i=p\;,\\
\quad -1 &\quad\mathrm{if}\,i=q\;.
\end{cases}\end{equation}
The matrices $\boldsymbol{G^c}$ and $\boldsymbol{G^a}$ have dimension $N\times N(N-1)/2$ and their expressions are, respectively, $g^c_{i\alpha}=f_{i\alpha}\sigma^c w^R_\alpha$ and $g^a_{i\alpha}=f_{i\alpha}\sigma^a w^R_\alpha$.
The trajectories obtained from the integration of the Langevin Equation~\eqref{eq:langevin_ehdpd} can be equivalently described as the evolution of a probability distribution for the state variables $\boldsymbol{y}$ obeying a Fokker-Planck equation~\cite{lau2007state,gubbiotti2019confinement}.
With the definitions~\eqref{y-def}--\eqref{g-def}
the Fokker-Planck equation associated with Eq.~\eqref{eq:langevin_ehdpd} reads
% FOKKER-PLANCK EHDPD
\begin{equation}
\label{eq:fokker-planck_ehdpd}
\frac{\partial P(\boldsymbol{y},t)}{\partial t}=\boldsymbol{\nabla}_y\cdot\left[\left(-\boldsymbol{u}+\frac{1}{2}\boldsymbol{\nabla}_y \cdot\boldsymbol{GG}^T+\frac{1}{2}\boldsymbol{GG}^T \cdot \boldsymbol{\nabla}_y \right)P(\boldsymbol{y})\right]\;,
\end{equation}
where $\boldsymbol{\nabla}_y = \left(\boldsymbol{\nabla}_x, \boldsymbol{\nabla}_v, \boldsymbol{\nabla}_{c}, \boldsymbol{\nabla}_{a} \right)^T$ is the 8N-dimensional gradient built with the derivatives with respect to the components of $\boldsymbol{y}$, Eq.~\eqref{y-def}.

It is convenient to introduce the matrix $\boldsymbol{D}=\boldsymbol{GG}^T/2$, which has dimension $8N\times 8N$ and can be decomposed in blocks
% D MATRIX
\begin{equation}
\boldsymbol{D}=\begin{pmatrix}
\boldsymbol{0} & \boldsymbol{0} & \boldsymbol{0} & \boldsymbol{0} \\
\boldsymbol{0} & \boldsymbol{D^v} & \boldsymbol{0} & \boldsymbol{0} \\
\boldsymbol{0} & \boldsymbol{0} & \boldsymbol{D^c} & \boldsymbol{0} \\
\boldsymbol{0} & \boldsymbol{0} & \boldsymbol{0} & \boldsymbol{D^a}
\end{pmatrix}\;,\end{equation}
% D BLOCKS
where $\boldsymbol{D}^{\boldsymbol v}$ is a $3N\times 3N$ submatrix, and both $\boldsymbol{D^c}$ and $\boldsymbol{D^a}$ are $N\times N$ submatrices.
The matrix $\boldsymbol{D}^{\boldsymbol v}$ can be further decomposed in $N\times N$ blocks of $3\times3$ submatrices $\boldsymbol{D}_{ij}^v$.
Their expressions are as follows 
\begin{equation}\label{eq:d_blocks}\begin{aligned}
\boldsymbol{D}_{ij}^v=&
\frac{1}{2} \sum\limits_{\alpha }  \boldsymbol{g}^v_{i\alpha}\otimes\boldsymbol{g}^v_{j\alpha}=\frac{1}{2m^2}\sum\limits_{\alpha }   f_{i\alpha}f_{j\alpha}\left(\sigma w^R_\alpha\right)^2\boldsymbol{e}_\alpha\otimes\boldsymbol{e}_\alpha\;,\\
D^c_{ij}=&\frac{1}{2} \sum\limits_{\alpha } g^c_{i\alpha}g^c_{j\alpha}=\frac{1}{2} \sum\limits_{\alpha }   f_{i\alpha}f_{j\alpha}\left(\sigma^c w^R_\alpha\right)^2\;,\\
D^a_{ij}=&\frac{1}{2} \sum\limits_{\alpha } g^a_{i\alpha}g^a_{j\alpha}=\frac{1}{2} \sum\limits_{\alpha }  f_{i\alpha}f_{j\alpha}\left(\sigma^a w^R_\alpha\right)^2\;,
\end{aligned}\end{equation}
where the summation over $\alpha$ between particle  pairs is explicitly indicated.
These expressions can be simplified accounting for the properties of product $f_{i\alpha}f_{j\alpha}$, i.e. $f_{i\alpha}f_{j\alpha} = 0$ except  two special cases: i) $f_{i\alpha}f_{j\alpha} = 1$ for $i = j$ and $\alpha = \alpha(i,q)$ or $\alpha = \alpha(p,i)$ for any $p$ and $q$; ii) $f_{i\alpha}f_{j\alpha} = -1$ for $i \ne j$ and $\alpha = \alpha(i,j)$ or $\alpha = \alpha(j,i)$.

With this in mind, Eq.~\eqref{eq:d_blocks} is rewritten as
% D BLOCKS SIMPLIFIED
\begin{align}
\label{eq:d_blocks_v}
\boldsymbol{D}^v_{ij}=&
\frac{1}{2m^2}\delta_{ij}\sum\limits_{k\ne i}\left(\sigma w^R_{ik}\right)^2\boldsymbol{e}_{ik}\otimes\boldsymbol{e}_{ik}+\frac{1}{2m^2}\left(\delta_{ij}-1\right)\left(\sigma w^R_{ij}\right)^2\boldsymbol{e}_{ij}\otimes\boldsymbol{e}_{ij}\;,\\
\label{eq:d_blocks_c}
D^c_{ij}=&\frac{1}{2}\delta_{ij}\sum\limits_{k\ne i}\left(\sigma^c w^R_{ik}\right)^2+\frac{1}{2}\left(\delta_{ij}-1\right)\left(\sigma^c w^R_{ij}\right)^2\;,\\
\label{eq:d_blocks_a}
D^a_{ij}=&\frac{1}{2}\delta_{ij}\sum\limits_{k\ne i}\left(\sigma^a w^R_{ik}\right)^2+\frac{1}{2}\left(\delta_{ij}-1\right)\left(\sigma^a w^R_{ij}\right)^2\; .
\end{align}
In the following it is assumed that the dynamics generated by Eq.~\eqref{eq:langevin_ehdpd} admits an equilibrium distribution
% BOLTZMANN
\begin{equation}
\label{eq:boltzmann_eq}
P_{eq}(\boldsymbol{y}) = C \exp\left[S(\boldsymbol{y})/k_B\right]\; ,
\end{equation}
where $S(\boldsymbol{y}) = k_B (\ln \left[ P_{eq}(\boldsymbol{y}) \right] + \ln C)$ is an appropriate thermodynamic potential depending on the coarse-grained variables 
$\boldsymbol{y}$ and $k_B$ is Boltzmann constant. In the present context, dealing with an isolated system, S can be understood as the (coarse-grained) entropy of the system. 
The equilibrium distribution must be the solution of the Fokker-Planck Eq.~\eqref{eq:fokker-planck_ehdpd}, i.e. 
% EQUILIBRIUM FOKKER-PLANCK EHDPD
\begin{equation}
\label{eq:fokker-planck_equilibrium}
0=\boldsymbol{\nabla}_y \cdot\left[\left(-\boldsymbol{u}+\boldsymbol{\nabla}_y \cdot\boldsymbol{D}+\boldsymbol{D} \cdot \boldsymbol{\nabla}_y S\right)P_{eq}(\boldsymbol{y})\right]\; .
\end{equation}
For the following calculations is convenient to split the drift term $\boldsymbol{u}$ into two parts, the conservative drift
% CONSERVATIVE DRIFT
\begin{equation}
\label{eq:conservative_drift}
\boldsymbol{u}^{\boldsymbol{C}}=\begin{pmatrix}
\boldsymbol{v}_1\\
\ldots \\
\boldsymbol{v}_N\\
m^{-1}\boldsymbol{f}^{\boldsymbol{C}}_{1}\\
\ldots \\
m^{-1}\boldsymbol{f}^{\boldsymbol{C}}_{N}\\
\boldsymbol{0}\\
\boldsymbol{0}
\end{pmatrix}\;,\end{equation} 
and the dissipative drift, 
% DISSIPATIVE DRIFT
\begin{equation}
\label{eq:dissipative_drift}
\boldsymbol{u}^{\boldsymbol{D}}=\begin{pmatrix}
\boldsymbol{0}\\
m^{-1}\sum\limits_{j\ne 1}
\gamma w^D_{1j}\left(\boldsymbol{v}_{j1}\cdot\boldsymbol{e}_{1j}\right)\boldsymbol{e}_{1j}\\
\ldots\\
m^{-1}\sum\limits_{j\ne N}
\gamma w^D_{Nj}\left(\boldsymbol{v}_{jN}\cdot\boldsymbol{e}_{Nj}\right)\boldsymbol{e}_{Nj}\\
\sum\limits_{j\ne 1}
\gamma^cw^D_{1j}h_{1j}^c\\
\ldots \\
\sum\limits_{j\ne N}
\gamma^cw^D_{Nj}h_{Nj}^c\\
\sum\limits_{j\ne 1}
\gamma^aw^D_{1j}h_{1j}^a \\
\ldots \\
\sum\limits_{j\ne N}
\gamma^aw^D_{Nj}h_{Nj}^a
\end{pmatrix}\;.\end{equation}
such that $\boldsymbol{u}=\boldsymbol{u^C}+\boldsymbol{u^D}$.

\subsection{FDR for hydrodynamics}

The Fluctuation Dissipation Relation (FDR) for pure hydrodynamics is discussed there to recover within the present framework the classical DPD expression.
The requirement that the conservative interactions alone should leave the equilibrium distribution unchanged implies that
% LIOUVILLE
\begin{equation}
\label{eq:Conservative}
\boldsymbol{\nabla}_y\cdot\left(\boldsymbol{u^C}P_{eq}\right)=
\boldsymbol{u^C}\cdot\boldsymbol{\nabla}_yP_{eq}=
\left[\boldsymbol{u^C}\cdot\boldsymbol{\nabla}_y\left(\frac{S}{k_B}\right)\right]P_{eq}=0\;,
\end{equation}
where $\boldsymbol{\nabla}_y \cdot \boldsymbol{u^C} = 0$ and $\boldsymbol{\nabla}_y P_{eq} = P_{eq} \boldsymbol{\nabla}_y S/k_B$.
This condition is satisfied if
\begin{align}
\label{eq:cond_S_1}\frac{\partial S}{\partial \boldsymbol{x}_i}\propto\boldsymbol{f}^{\boldsymbol C}_i\;,\\
\label{eq:cond_S_2}\frac{\partial S}{\partial \boldsymbol{v}_i}\propto - m\boldsymbol{v}_i\;.
\end{align}
As will be clear from the following, the proportionality constant is assumed to be $1/T$, with $T$ the temperature, taken to be uniform throughout the system. Based on Eq.~\eqref{eq:Conservative} and Eq.~\eqref{eq:fokker-planck_equilibrium} the expression
% DRIFT EXPRESSION
\begin{equation}
\label{eq:drift_expression}
\boldsymbol{u^D}=\boldsymbol{\nabla}_y\cdot\boldsymbol{D}+\boldsymbol{D} \cdot \boldsymbol{\nabla}_yS\;
\end{equation}
gives a condition for the Fokker-Planck equation~\eqref{eq:fokker-planck_ehdpd} to have the equilibrium solution of Eq.~\eqref{eq:boltzmann_eq} with $S$ satisfying Eq.~\eqref{eq:cond_S_1} and Eq.~\eqref{eq:cond_S_2}.
The Fokker-Planck equation can then be rewritten as follows
\begin{equation}\label{eq:fokkerplanck}
\dot{P}=-\boldsymbol{\nabla}\cdot\left[\left(\boldsymbol{u}^C+\boldsymbol{D}\cdot\boldsymbol{\nabla}S-\boldsymbol{D}\cdot\boldsymbol{\nabla}\right)P\right]\;.
\end{equation}
Using Eq,~\eqref{eq:dissipative_drift}, as well as the definition of dissipative drift Eq.~\eqref{eq:drift_expression}, gives for the velocity component of meso-particle $i$,
% DRIFT VELOCITY
\begin{equation}
\label{eq:drift_velocity}
m^{-1}\sum\limits_{j\ne  i}
\gamma w^D_{ij}\left(\boldsymbol{v}_{ji}\cdot\boldsymbol{e}_{ij}\right)\boldsymbol{e}_{ij}=
\sum\limits_{j\ne i}
\frac{1}{2m^2}\left(\sigma w^R_{ij}\right)^2\boldsymbol{e}_{ij}\otimes\boldsymbol{e}_{ij}\frac{1}{k_B}\left(\frac{\partial S}{\partial\boldsymbol{v}_i}-\frac{\partial S}{\partial \boldsymbol{v}_j}\right)\;,
\end{equation}
where we used the fact that the matrix $\boldsymbol{D}$ doesn't depend on the velocity, and hence $\boldsymbol{\nabla}_v\cdot\boldsymbol{D}=0$.
Considering Eq.~\eqref{eq:cond_S_2}, Eq.~\eqref{eq:drift_velocity} is identically satisfied if the following two conditions are met
% FLUCTUATION DISSIPATION VELOCITY
\begin{equation}\label{eq:fluctuation-dissipation_velocity}\begin{aligned}
\gamma=\frac{\beta\sigma^2}{2}\;,\\
w^D_{ij}=\left(w^R_{ij}\right)^2
\;,\end{aligned}\end{equation}
with $\beta^{-1}= k_B T$.
These are the same FDRs  found by Espa\~nol and Warren~\cite{espanol1995statistical} for hydrodynamics. 
\subsection{FDR for Ionic transport}
The Fluctuation Dissipation Relations (FDRs) discussed in the previous sections for pure hydrodynamics are here extended to include ionic transport. Clearly all the results previously established still hold in presence of transported charges. 

It is instrumental to introduce the problem by considering the simplest case of constant noise intensities $\sigma^c$ and $\sigma^a$ which imply $\boldsymbol{\nabla}_c\cdot\boldsymbol{D}^c=0$ and $\boldsymbol{\nabla}_a\cdot\boldsymbol{D}^a=0$.
Using Eqs.~\eqref{eq:dissipative_drift} and~\eqref{eq:drift_expression}, and using the expressions for $\boldsymbol{D}^c$ and $\boldsymbol{D}^a$ from Eq.~\eqref{eq:d_blocks_c} and Eq.~\eqref{eq:d_blocks_a}, we obtain
% DRIFT CHARGE 1
\begin{equation}\label{eq:drift_charge_1}\begin{aligned}
\sum\limits_{j\ne i}
\gamma^cw^D_{ij}h_{ij}^c=
\frac{1}{2}\left(\sigma^c w^R_{ij}\right)^2\frac{1}{k_B}\left(\frac{\partial S}{\partial n^c_i}-\frac{\partial S}{\partial n^c_j}\right)\;,\\
\sum\limits_{j\ne i}
\gamma^aw^D_{ij}h_{ij}^a=
\frac{1}{2}\left(\sigma^a w^R_{ij}\right)^2\frac{1}{k_B}\left(\frac{\partial S}{\partial n^a_i}-\frac{\partial S}{\partial n^a_j}\right)
\;.\end{aligned}\end{equation}
Exploiting the second condition in \eqref{eq:fluctuation-dissipation_velocity} for the weight functions, we obtain % after defining the chemical potential of particle $i$ as 
\begin{equation}\label{eq:fluctuation-dissipation_charge}\begin{aligned}
&\gamma^c=\frac{\beta}{2} \sigma_c^2\;,\\
&\gamma^a=\frac{\beta}{2} \sigma_a^2
\;\end{aligned}\end{equation}
and 
\begin{equation}\begin{aligned}
h^c_{ij}=&T\left(\frac{\partial S}{\partial n^c_i}-\frac{\partial S}{\partial n^c_j}\right)=\mu^c_j-\mu^c_i\;,\\
h^a_{ij}=&T\left(\frac{\partial S}{\partial n^a_i}-\frac{\partial S}{\partial n^a_j}\right)=\mu^a_j-\mu^a_i
\;,\end{aligned}\end{equation}
where the chemical potentials of the meso-particle are defined as
\begin{equation}\begin{aligned}
\label{eq:chemical_S}
\mu^c_i=&-T\frac{\partial S}{\partial n^c_i}\;,\\
\mu^a_i=&-T\frac{\partial S}{\partial n^a_i}\;.
\end{aligned}\end{equation}

% \subsection{FDR for Ionic transport: the general case}
%

Removing the assumption of constant noise intensity, $\sigma^c$ and $\sigma^a$ can now depend on cations and anions carried by the interacting particles, i.e. $\sigma^c=\sigma^c(n^c_i,n^c_j)=\sigma^c_{ij}$ and $\sigma^a=\sigma^a(n^a_i,n^a_j)=\sigma^a_{ij}$ with $\sigma^c_{ij}=\sigma^c_{ji}$ and $\sigma^a_{ij}=\sigma^a_{ji}$.
Hence $\boldsymbol{\nabla}_c\cdot\boldsymbol{D}^c$ and $\boldsymbol{\nabla}_a\cdot\boldsymbol{D}^a$ do not vanish, in general, and must be considered in the drift term, Eq.~\eqref{eq:drift_expression}.
From the expressions for $\boldsymbol{D}^c$ and $\boldsymbol{D}^a$, Eq.~\eqref{eq:d_blocks_c} and Eq.~\eqref{eq:d_blocks_a}, the dissipative drifts now read
% DRIFT CHARGE 1
\begin{equation}\label{eq:drift_charge_1bis}\begin{aligned}
\sum\limits_{j\ne i}
\gamma^cw^D_{ij}h_{ij}^c=
\frac{1}{2}\sum\limits_{j\ne i}\left(\sigma^c_{ij} w^R_{ij}\right)^2\left[\frac{1}{\left(\sigma^c_{ij}\right)^2}\left(\frac{\partial \left(\sigma^c_{ij}\right)^2}{\partial n^c_i}-\frac{\partial \left(\sigma^c_{ij}\right)^2}{\partial n^c_j}\right)+
\frac{1}{k_B}\left(\frac{\partial S}{\partial n^c_i}-\frac{\partial S}{\partial n^c_j}\right)\right]\;,\\
\sum\limits_{j\ne i}
\gamma^aw^D_{ij}h_{ij}^a=
\frac{1}{2}\sum\limits_{j\ne i}\left(\sigma^a_{ij} w^R_{ij}\right)^2\left[\frac{1}{\left(\sigma^a_{ij}\right)^2}\left(\frac{\partial \left(\sigma^a_{ij}\right)^2}{\partial n^a_i}-\frac{\partial \left(\sigma^a_{ij}\right)^2}{\partial n^a_j}\right)+
\frac{1}{k_B}\left(\frac{\partial S}{\partial n^a_i}-\frac{\partial S}{\partial n^a_j}\right)\right]\;,\\
\;\end{aligned}\end{equation}
which, using  Eq.~\eqref{eq:fluctuation-dissipation_velocity} for the weight functions together with Eq.~\eqref{eq:fluctuation-dissipation_charge} and the definitions \eqref{eq:chemical_S}, gives the explicit expression for $h^c_{ij}$ and $h^a_{ij}$ in the general case
\begin{equation}\label{eq:h}\begin{aligned}
h^c_{ij}=\frac{1}{\beta\gamma^c_{ij}}\left(\frac{\partial \gamma^c_{ij}}{\partial n^c_i}-\frac{\partial\gamma^c_{ij}}{\partial n^c_j}\right)+\mu^c_{ji}\;,\\
h^a_{ij}=\frac{1}{\beta\gamma^a_{ij}}\left(\frac{\partial \gamma^a_{ij}}{\partial n^a_i}-\frac{\partial\gamma^a_{ij}}{\partial n^a_j}\right)+\mu^a_{ji}\;.
\end{aligned}\end{equation}
It is worth noting that  the antisymmetry conditions~\eqref{eq:h_antisymm} are satisfied, implying conservation of the total quantity of each ionic species.
As typical, when the noise intensity depends on the state variables, additional drift terms appear in the Langevin equation, see also \cite{lau2007state,gubbiotti2019confinement}.

\section{Physical model}
\label{sec:model}

In the previous section it was shown that if the conditions~\eqref{eq:fluctuation-dissipation_charge}, \eqref{eq:fluctuation-dissipation_velocity}, \eqref{eq:cond_S_1} and \eqref{eq:cond_S_2} are met, system~(\ref{eq:ehdpd_1}-\ref{eq:ehdpd_4}) admits a stationary equilibrium solution given by~\eqref{eq:boltzmann_eq}. The conservative force $\boldsymbol{f}^C_i$, the particle velocity $\boldsymbol{v}_i$ and the chemical potentials $\mu^c_i$ and $\mu^a_i$ are related to the entropy of the system $S$ through Eqs.~\eqref{eq:cond_S_1}, \eqref{eq:cond_S_2} and \eqref{eq:chemical_S}.
The system (total) energy $E_{TOT}$ may be expressed as
\begin{equation}\label{eq:toteng}
E_{TOT}=U_E+\sum\limits_{i=1}^N\left(\frac{m}{2}\boldsymbol{v}_i^2+U_i\right)\;,
\end{equation}
where $U_E$ is the electrostatic energy of the system. 
The sum includes the portion of kinetic energy and internal energy 
associated with particle $i$, $m\boldsymbol{v}^2_i/2$  and $U_i$, respectively.
Under the assumption of local equilibrium, the particle entropy can be expressed in terms of Helmholtz free energy $A_i$ and internal energy, 
\begin{equation}
S_i=\frac{1}{T}\left(U_i-A_i\right)\;.
\end{equation}
Hence, the total entropy reads
\begin{equation}\label{eq:entropy}
S=\sum\limits_{i=1}^NS_i=\sum\limits_{i=1}^N\frac{1}{T}\left(U_i-A_i\right)=\frac{1}{T}\left[E_{TOT}-U_{E}-\sum\limits_{i=0}^N \left(A_i+\frac{m}{2}\boldsymbol{v}^2_i\right)\right]\; .
\end{equation}
For an isolated system $E_{TOT} = const$ and the entropy $S$ is fully specified once free energy $A_i$ and electrostatic energy $U_E$ are given in terms of the coarse-grained variables.
In general terms, the Helmholtz free energy density (per unit mass) is a function of specific volume, temperature and number densities. As a consequence, the
free energy of the meso-particle depends on particle volume $V_i$, (uniform) temperature, and composition given in terms of number of atoms (in the sense of indivisible particles) belonging to the meso-particle. In the following the composition is specified in terms of $n^s_i$, $n^c_i$, $n^a_i$, the number of solvent, cationic and anionic atoms, respectively. Hence,  $A_i = A_i(V_i,T,n_i^s, n_i^c, n_i^a)$.

Before specifying the free energy in detail, it is worth defining first the meso-particle volume,
\begin{equation}\label{eq:density}
V_i^{-1}=\sum\limits_{j=1}^Nw(r_{ij})\;,
\end{equation}
where $w(r)$ is a  differentiable  compactly supported, positive function with (single) maximum at $r=0$, with integral normalized to $1$, that vanishes identically
for $r$ larger than a cutoff $r_c$.
%Here, the function $w^D$ as defined in Eq.~\eqref{eq:weight} is used $w(r)=w_0w^D(r)$, where $w_0=15/2\pi$ is a normalization factor.
Hereafter, $w(r)$ is specified as the Lucy's function commonly used in the context of Smoothed Dissipative Particle Dynamics~\cite{espanol2003smoothed},
\begin{equation}
w\left({r}/{r_c}\right)=\begin{cases}
\frac{\displaystyle 105}{\displaystyle 16 \pi \, r_c^3}\left(1+3r/r_c\right)\left(1-r/r_c\right)^3\quad\mathrm{if}\quad r/r_c<1\;,
\\0\quad\mathrm{if}\quad r/r_c>1\; .
\end{cases}
\end{equation}

	%%%%%%%%%%%%%%%%%%%%%%
	% CHEMICAL POTENTIAL %
	%%%%%%%%%%%%%%%%%%%%%%

\subsection{Free energy}

As partially anticipated, the meso-particle is considered to be constituted by $M$ atoms of three different kinds and equal mass,
namely $n^c_i$ cations, $n^a_i$  anions and  $n^s_i$ solvent atoms.
Number of cations, anions and solvent atoms may change during the dynamics under the constraint of constant total mass,
\begin{equation}\label{eq:constant}
n^c_i+n^a_i+n^s_i=M=\mathrm{const}\;.
\end{equation}
A simple model for the particle free energy $A_i$ follows by considering a system comprising three non interacting species (perfect gas model)~\cite{statistical_mechanics}
\begin{equation}
\label{eq:free-energy}
A_i=
\frac{n^c_i}{\beta}\left[\log\left(\frac{n^c_i\lambda^3}{V_i}\right)-1\right]
	+\frac{n^a_i}{\beta}\left[\log\left(\frac{n^a_i\lambda^3}{V_i}\right)-1\right]+\frac{n^s_i}{\beta}\left[\log\left(\frac{n^s_i\lambda^3}{V_i}\right)-1\right]\;,
\end{equation}
where $\lambda$, depending on temperature and atoms masses, is the De Broglie's thermal wavelength.
In Eq.~\eqref{eq:free-energy} the dependence on $n^s_i$ may eliminated in favor of the total number of atoms forming the meso-particle, Eq.~\eqref{eq:constant}, while the temperature, and hence $\lambda$, is assumed to be the same in all the meso-particles.

Using the constraint of Eq.~\eqref{eq:constant} to eliminate the number of solvent atoms $n_s$, Eq.~\eqref{eq:free-energy} reads
\begin{equation}\label{eq:free_energy}
	A_i(V_i,n^c_i,n^a_i)=\frac{n^c_i}{\beta}\left[\log\left(\frac{n^c_i}{M-n^c_i-n^a_i}\right)-1\right]+\frac{n^a_i}{\beta}\left[\log\left(\frac{n^a_i}{M-n^c_i-n^a_i}\right)-1\right]+\frac{M}{\beta}\log\left(\frac{M-n^a_i-n^c_i}{V_i}\right)\;,
\end{equation}
where inessential constant terms have been omitted.
Notice that the particle pressure, related to the derivative of the free energy with respect to volume, turns out to depend on the total number of atoms $M$, 
see \S~\ref{sec:conservative} below.

	%%%%%%%%%%%%%%%%%%
	% ELECTROSTATICS %
	%%%%%%%%%%%%%%%%%%

\subsection{Electrostatics of EH-DPD particles}

The expression~\eqref{eq:entropy} for the entropy requires an explicit form for the electrostatic energy.
The coarse-grained variables $n^c_i$ and $n^a_i$ provide the number of cations and anions carried by the meso-particle. Its charge is then 
\begin{equation}\label{eq:charge}
q_i=q_cn^c_i-q_an^a_i\;,
\end{equation}
with $q_c$ and $q_a$  the (absolute value of the) charge of a single cation/anion.
The charge distribution associated to each meso-particle is given by as a Gaussian function centered in $\boldsymbol{x}_i$  with constant variance $s^2$, i.e. 
% GAUSSIAN 
\begin{equation}\label{eq:gaussian}
\rho_i(\boldsymbol{r})=\rho(\boldsymbol{r},\boldsymbol{x}_i,q_i)=\frac{q_i}{\left(2\pi s^2\right)^{3/2}}\exp{\frac{-|\boldsymbol{r}-\boldsymbol{x}_i|^2}{2s^2}}\; .
\end{equation}
The energy ascribed to the interaction between meso-particles, $i\ne j$, is then~\cite{kiss2014efficient}
% PAIR ENERGY
\begin{equation}
U^E_{ij} = U^E_{ji} = \frac{1}{2}
\int\int\frac{\rho_i(\boldsymbol{r})\rho_j(\boldsymbol{r'})}{|\boldsymbol{r}-\boldsymbol{r'}|}\mathrm{d}\boldsymbol{r}\mathrm{d}\boldsymbol{r'}
=\frac{q_iq_j}{r_{ij}}\erf\left(\frac{r_{ij}}{2s}\right)\;,
\end{equation}
where the interaction energy of the couple is $U^E_{ij} + U^E_{ji}$.
Introducing  the self-energy, $i=j$, 
% SELF ENERGY
\begin{equation}
U^E_{ii}=
\frac{1}{2}\int\int\frac{\rho_i(\boldsymbol{r})\rho_i(\boldsymbol{r'})}{|\boldsymbol{r}-\boldsymbol{r'}|}\mathrm{d}\boldsymbol{r}\mathrm{d}\boldsymbol{r'}
=\frac{q^2_i}{2s\sqrt{\pi}}\;.
\end{equation}
The self-energy does not contribute to the electrostatic force, since it is independent of the relative positions of the meso-particles. However, it does contribute to the total electrostatic potential of the meso-particle, which can be defined as 
% ELECTROSTATIC POTENTIAL
\begin{equation}\label{eq:electrostatic_potential}
\Phi_i=\frac{\partial U^E}{\partial q_i}=\frac{q_i}{s\sqrt\pi}+\sum\limits_{i\ne j}^N \frac{q_j}{r_{ij}}\erf\left(\frac{r_{ij}}{2s}\right)\;.
\end{equation}
Since $\lim\limits_{r\to 0}\erf(r/(2s))/r=1/(s\sqrt\pi)$, the above expression can be rewritten in compact form as
\begin{equation}\label{eq:particle_potential}
\Phi_i=\sum\limits_{j=1}^N \frac{q_j}{r_{ij}}\erf\left(\frac{r_{ij}}{2s}\right)\;,
\end{equation}
where now the summation also includes the term $j=i$ ($r_{ii}=0$).
Finally, the total electrostatic energy of the system can be expressed as
% ELECTROSTATIC ENERGY PARTICLE
\begin{equation}\label{eq:electrostatic_energy}
U_E=\frac{1}{2}\sum\limits_{i=1}^Nq_i\Phi_i \; .
\end{equation}

\subsection{Chemical potential and conservative force}
\label{sec:conservative}
Specifying the electrostatic energy completes the expression of the entropy, Eqs.~\eqref{eq:entropy}, providing the chemical potential, Eqs.~\eqref{eq:chemical_S}. Its explicit expression for cations and anions is
\begin{equation}\begin{aligned}\label{eq:chemical}
\mu^c_i=-T\frac{\partial S}{\partial n^c_i}=\frac{\partial \left(A_i+U_E\right)}{\partial n^c_i}=\frac{1}{\beta}\log\left(\frac{n^c}{M-n^c-n^a}\right)+q_c\Phi_i\;,\\
\mu^a_i=-T\frac{\partial S}{\partial n^a_i}=\frac{\partial \left(A_i+U_E\right)}{\partial n^a_i}=\frac{1}{\beta}\log\left(\frac{n^a}{M-n^c-n^a}\right)-q_a\Phi_i\;,\\
\end{aligned}\end{equation}
respectively, where constant terms have been omitted, since the dynamics depends only on the chemical potential differences.
The effect of electrostatic interactions,  proportional to the particle electrostatic potential, adds to the familiar contribution coming from the (perfect gas) equation of state. 

The conservative component of the force acting on the particle is obtained by differentiating the entropy $S$ with respect to particle position, Eq.~\eqref{eq:cond_S_1},
\begin{equation}
\boldsymbol{f}^{\boldsymbol {C}}_i=T\frac{\partial S}{\partial\boldsymbol{x}_i}=-\frac{\partial\left(U_E+A\right)}{\partial\boldsymbol{x}_i}=\boldsymbol{f}^{\boldsymbol {E}}_i+\boldsymbol{f}^{\boldsymbol {P}}_i\;,
\end{equation}
where $A=\sum\limits_{j=1}^NA_j$ is the system free energy. As for the chemical potential, the conservative force comes form two contributions. The origin of the electrostatic force $\boldsymbol{f}^{\boldsymbol {E}}_i$  is immediately clear.  It can be computed using Eqs.~\eqref{eq:electrostatic_energy} and \eqref{eq:electrostatic_potential}, giving
\begin{equation}
\boldsymbol{f}^{\boldsymbol{E}}_i=-\sum\limits_{j\ne i}q_i\frac{\partial\Phi_i}{\partial r_{ij}}\boldsymbol{e}_{ij}=\sum\limits_{j\ne i}q_iq_j\frac{\sqrt\pi s\erf\left(r_{ij}/(2s)\right)-r_{ij}\exp\left(-r_{ij}^2/(4s^2)\right)}{s\sqrt\pi r^2_{ij}}\boldsymbol{e}_{ij}\;.
\end{equation}
The second contribution follows from the equation for the free energy $A$, \eqref{eq:free_energy}, giving
\begin{equation}\label{eq:pforce}
\boldsymbol{f}^{\boldsymbol{P}}_i=-\sum\limits_{j\ne i}\frac{\partial A_j}{\partial{V_j}}\frac{\partial V_j}{\partial \boldsymbol{x}_i}=\sum\limits_{j\ne i}\frac{M}{\beta}\left(V_j+V_i\right)w'_{ij}\boldsymbol{e}_{ij}\;,
\end{equation}
where Eq.~\eqref{eq:density} has been used and $w'_{ij}$ is the derivative of the weight function $w(r_{ij})$. It could be noted that $- \partial A_i/\partial V_i = M/\beta V_i$ is the meso-particle pressure $p_i$, providing the standard interpretation of $\boldsymbol{f}^{\boldsymbol{P}}_i$ as the pressure force. 

	%%%%%%%%%%%%%%%%
	% IONIC FLUXES %
	%%%%%%%%%%%%%%%%

\begin{figure}
\includegraphics[width=0.5\linewidth]{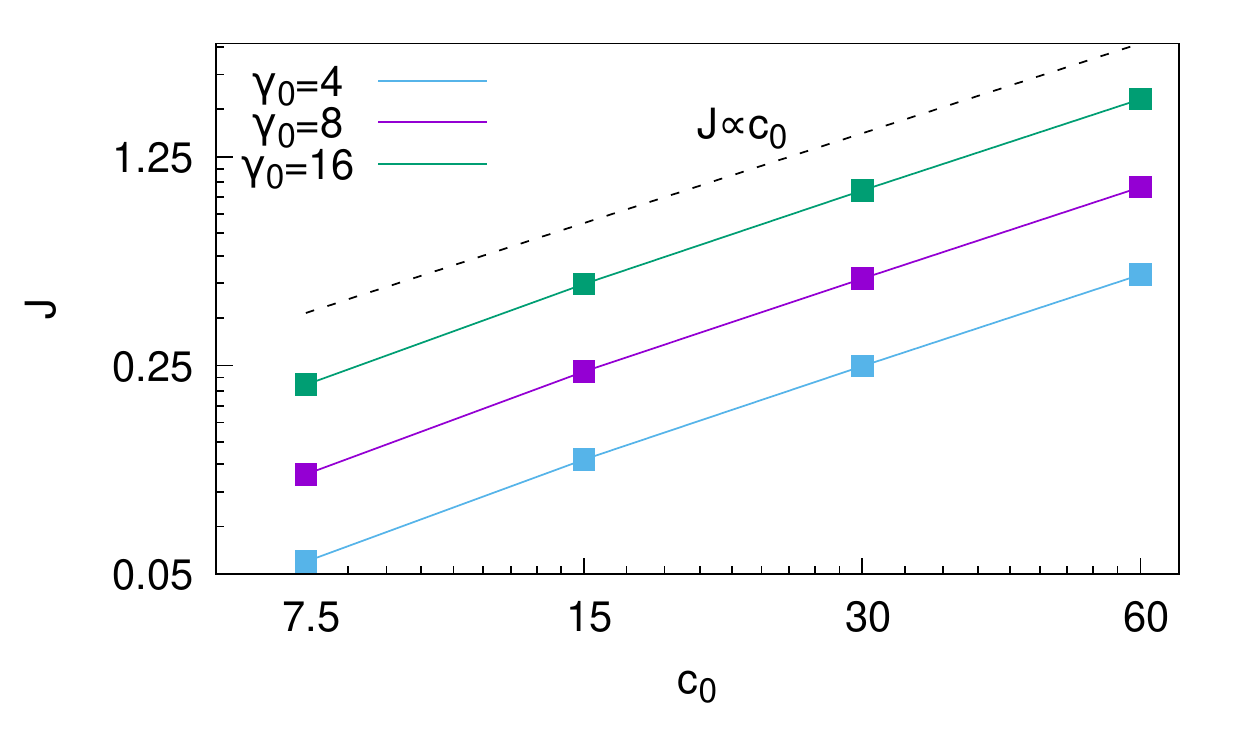}
\caption{ \label{fig:cond2}
	Ionic current density $J$ as defined in Appendix~\ref{sec:current} resulting after the application of an electric field $E=10$, in the direction parallel to the application of the field. The current density was computed for different values of the dissipative parameters $\gamma^a_0=\gamma^c_0=\gamma_0$ and of average concentration $c_0$. The electric current density (and hence the conductivity) shows a linear dependence on the ion concentration ${c_0}$ for the range of concentrations simulated. For these simulations a value of $q^c=q^a=q=0.03635$ has been used.
}
\end{figure}

\subsection{Ionic exchange between particles}
\label{sec:dissipative_factors}

The coefficients $\gamma^{c/a}$ control the ionic exchange between particles and don't affect the equilibrium distribution. 
Instead, they control the cationic and anionic conductivity of the solution, see Appendix~\ref{sec:conductivity}.
The following expression was chosen for the coefficients $\gamma^{c/a}$
\begin{equation}\begin{aligned}\label{eq:gamma0}
&\gamma_c(n^c_i,n^c_j)=\gamma^c_0\sqrt{n^c_in^c_j}\\
&\gamma_a(n^a_i,n^a_j)=\gamma^a_0\sqrt{n^a_in^a_j}\;,
\end{aligned}\end{equation}
where $\gamma^{c/a}_0$ are parameters which control the ionic current between particles at a given concentration.
The resulting ionic conductivity shows a linear dependence on the ionic concentration ${c_0}$ and on the parameters $\gamma^{c/a}_0$, at least in the explored range of parameters, see Fig.~\ref{fig:cond2}.
As previously shown, Eqs.~\eqref{eq:h}, the use of parameters $\gamma^{c/a}$ which depend on concentration gives rise to an additional exchange of ions between the particles independent on the chemical potential
\begin{equation}\begin{aligned}
	&h^c_{ij}=\frac{1}{\beta}\left(\frac{1}{n^c_i}-\frac{1}{n^c_j}\right)+\mu^c_{ji}\\
	&h^a_{ij}=\frac{1}{\beta}\left(\frac{1}{n^a_i}-\frac{1}{n^a_j}\right)+\mu^a_{ji} \; .
\end{aligned}\end{equation}

\section{Validation}
\label{sec:validation}

As a case of study to validate the EH-DPD model, a system consisting of a planar channel of height $h$ with given surface charge at the two walls is simulated.
Before describing the actual set-up, it is instrumental to review classical solutions for electroosmotic flows
based on the Debye approximation $\beta\zeta q \ll 1$ where $\zeta$ is the wall
electric potential and $q=q_c=q_a$ is the ionic charge of the ions, where the electrolyte has been assumed symmetric.
In this case, the 1-D version of the Poisson-Boltzmann equation reads~\cite{theoretical_microfluidics}
\begin{equation}\label{eq:poissonboltzmann}
	\frac{\mathrm{d}^2\phi}{\mathrm{d}z^2}=
	\frac{2qc_0}{\varepsilon}\sinh(\beta q\phi(z)) \simeq \frac{\phi(z)}{\lambda_D^2}
	\;,
\end{equation}
where $z$ is the coordinate orthogonal to the channel walls, $\phi$ is the electrostatic potential, $c_0$ the concentration at zero potential and $
\varepsilon$ the dielectric constant.
In the linearized form on the right hand side of the equation, $\lambda_D=\left(2 \beta q^2c_0/\varepsilon\right)^{-1/2}$ is the Debye length.
The boundary conditions for this equation relate the derivative of the potential at the walls to the wall charge, i.e., assuming a vanishing electric field outside the channel
\begin{align}
	\frac{\mathrm{d}\phi}{\mathrm{d}z}\biggr\lvert_{z=h/2}=-\frac{\sigma_{up}}{\varepsilon}\;,
	\qquad\qquad\qquad\qquad
	\frac{\mathrm{d}\phi}{\mathrm{d}z}\biggr\lvert_{z=-h/2}=\frac{\sigma_{low}}{\varepsilon}\;,
\end{align}
where $\sigma_{up}$ and $\sigma_{low}$ are, respectively, the surface charges of the upper and lower walls.

We consider two different scenarios: i) the symmetric case (suffix $S$) with
both walls with same surface charge, $\sigma_{up}=\sigma_{low}=\sigma$ and ii) the antisymmetric case (suffix $A$)
where the two walls are oppositely charged, $\sigma_{up}=-\sigma_{low}=\sigma$.
In the two cases, the analytical solutions of Eq.~\eqref{eq:poissonboltzmann} read~\cite{micronanofluid}
\begin{align}
\phi^S=\zeta^S\frac{\cosh(z/\lambda_D)}{\cosh(h/(2\lambda_D))}\;,
	\qquad\qquad\qquad\qquad
\phi^A=\zeta^A\frac{\sinh(z/\lambda_D)}{\sinh(h/(2\lambda_D))}\;,
\label{eq:phi}
\end{align}
and the wall potentials ($\zeta$-potentials) in the symmetric and antisymmetric case are, respectively,
\begin{align}
	\zeta^S=\frac{\lambda_D\sigma}{\varepsilon\tanh(h/2\lambda_D)}\;,
		\qquad\qquad\qquad\qquad
	\zeta^A=\frac{\lambda_D\sigma\tanh(h/2\lambda_D)}{\varepsilon}\;.
\end{align}
The resulting cationic and anionic concentrations are $c^c(z)=c_0\exp\left(-q\phi\right)$ and $c^a(z)=c_0\exp\left(q\phi\right)$, respectively.
The electroosmotic velocity profile which arises after the application of an electric field $E$ parallel to the walls follows from the
Stokes equation endowed with no slip conditions at the walls,
\begin{align}
\eta \frac{d^2u}{dz^2} + qE (c^c - c^a) = 0\, ,
\end{align}
as
\begin{align}
u^S&=v_{eo}^S\left(1-\frac{\phi^S}{\zeta^S}\right)\;,\\
u^A&=v_{eo}^A\left(\frac{2z}{L}-\frac{\phi^A}{\zeta^A}\right)\;,
\end{align}
where $v_{eo}=\varepsilon \zeta E/\mu$ is the electroosmotic velocity and $\mu$ is the dynamic viscosity.

\subsection{Simulation set up}

For the validation simulations, we set equal to 1 the mass of the meso-particles $m$, the cutoff radius $r_c$ and the thermal energy $k_BT$. 
The remaining free parameters of the EH-DPD system are the number of atoms in the meso-particle $M$, the ion charge $q_a=q_c=q$, the parameter $s$ related to the Gaussian used to model electrostatic interactions, the dissipative coefficient $\gamma$ and the corresponding coefficients for the ionic transport $\gamma^c_0$ and $\gamma^a_0$ appearing in Eq.~(\ref{eq:gamma0}).
Setting $\gamma=1000$ leads to a viscosity of $\mu=86$ as computed by imposing a Poiseuille flow and measuring the resulting velocity profile~\cite{boromand2015viscosity}. 

The parameters $\gamma^c_0$ and $\gamma^a_0$ were calibrated, figure~\ref{fig:cond2},
 by estimating the electric conductivity obtained by applying a constant electric field to a triply-periodic EH-DPD system and measuring the resulting electric current 
 density as defined in Appendix~\ref{sec:current}.  
We did so for different values of  $\gamma^c_0$ and $\gamma^a_0$ and for different ionic concentrations $c_0$. 
We also estimated the conductivity of the fluid by an independent approach  based on linear response theory, see Appendix~\ref{sec:conductivity}, finding a good 
agreement with the nonequilibrium simulations.
Typical values used in the following subsection  are $\gamma^c_0 = \gamma^a_0 = 16$.

In the set of Eqs.~\eqref{eq:ehdpd_1} there is no guarantee that the quantities $n^c$ and $n^a$ are positive.
In fact, due to the stochastic nature of the equations, unfrequent, strong events can lead to a negative number of ions in the particle.
The chemical potential of Eq.~\eqref{eq:chemical} is not defined for negative $n^a_i$ and $n^c_i$, thus a limiting value of $\mu_{limit}=-10$ was used.
Also,  since $\gamma^a$ and $\gamma^c$ are not defined for negative $n^a_i$ and $n^c_i$, a lower cutoff  $\bar{n}$ is assumed  such that $n^a_i\ge\bar{n}$ and $n^c_i\ge\bar{n}$ (typically $\bar{n}=0.00223$ has been used in the simulations to be discussed).
The electrostatic interactions are dealt with the Ewald summation algorithm~\cite{kiss2014efficient} and the model was implemented using the DPD package of LAMMPS~\cite{lammps}. The Euler-Maruyama algorithm~\cite{higham2001algorithmic} was used to integrate Eq.~\eqref{eq:ehdpd_1} in the It\^o formalism using a time step of $\Delta t=10^{-5}$.

\subsection{Electroosmotic flow}

\begin{figure}
\centering\includegraphics[width=0.8\linewidth]{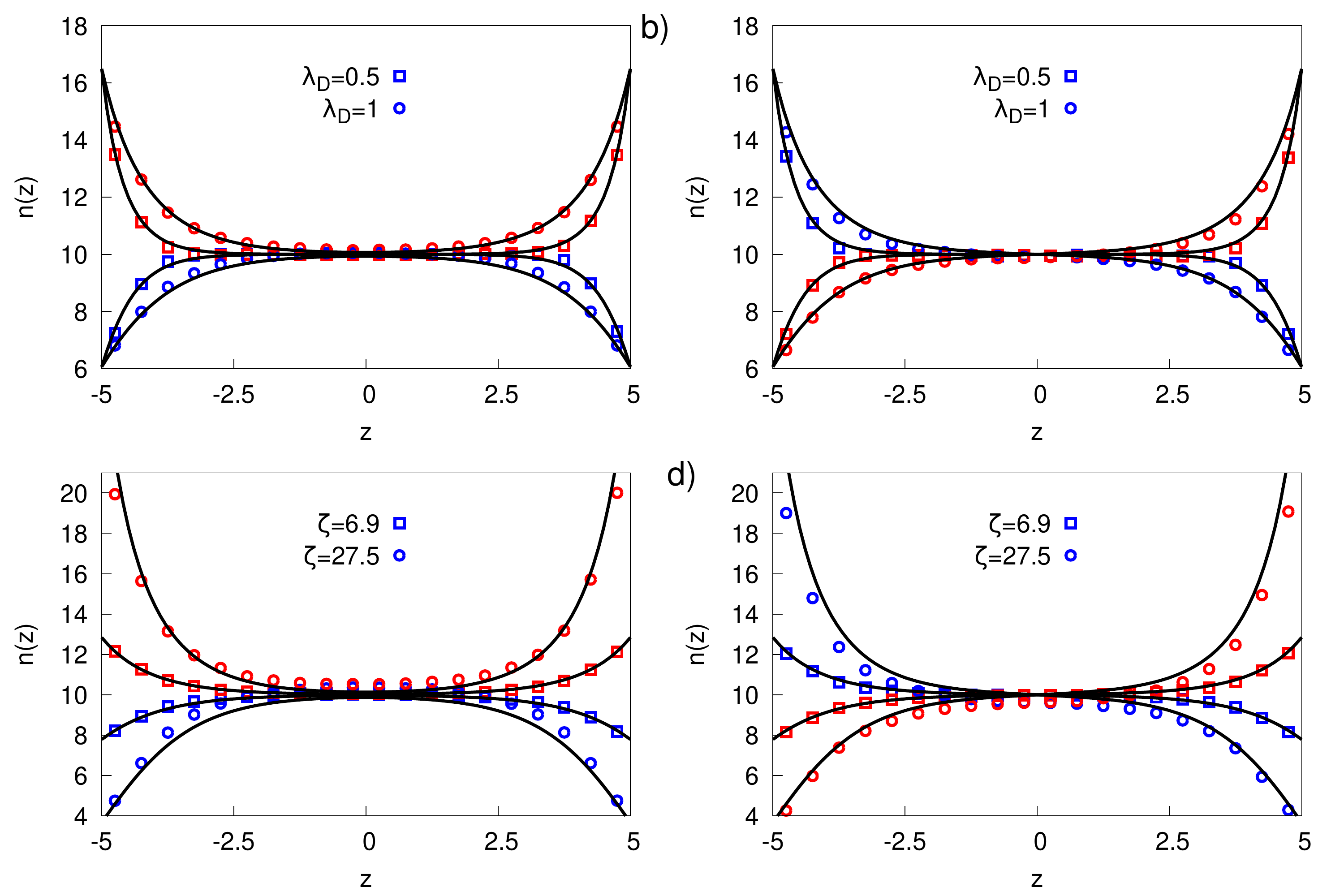}
\caption{
Ionic density profile for the cations (blue) and the anions (red) compared with the analytical solutions (black lines).
Figures a) and c) are cases with symmetric wall charges, while figures b) and d) are cases with antisymmetric wall charges.
In the two top plots, two different values of the Debye length are considered and $\zeta=13.7$, while in the bottom plots the zeta potential is changed and $\lambda_D=1$.
}
\label{fig:charge}
\end{figure}

We simulated a planar channel with a height of 10, ranging from $z=-5$ to $z=5$, where $z$ is the coordinate orthogonal to the walls. 
The walls were modelled using fixed meso-particles of constant charge distributed randomly in two layers of width 1 for
each side. The first layer, in direct contact with the fluid, has a particle number density of $\rho_1 = 3$, while the second layer, also of width 1, has a particle number density of $\rho_2 = 6$.
In a previous work~\cite{gubbiotti2019confinement}, we showed that, in the DPD context, walls constituted by fixed random particles of variable density are suitable to guarantee impermeability and a low slip while controlling density fluctuations due to fluid layering. 
A similar model was employed here the main difference being 
that now the wall is constituted by
two layers with different densities. 
The one exposed towards the liquid
mainly controls the slippage,
while the external one guarantees wall impermeability. 
The wall particles interact with the fluid particles via the multi-body 
potential which defines the pressure force, see Eq.~\eqref{eq:pforce}, 
where for the wall particle volume a constant value of $V_{wall}=0.8$
has been used for the inner layer and $V_{wall}=10$ for the external one.
In case the wall is charged, also the electrostatic forces are included.
We measured the slip length of the resulting fluid-wall system by imposing 
a Poiseuille flow, observing an acceptably low slip length $<2\%$ of the channel height.
The ionic charge $q$, considered equal for both species, has been tuned to adjust the Debye screening length, using $c_0=30$. The concentrations of the two ionic species are set to guarantee the global electroneutrality of the system, with the additional condition $\sqrt{c^cc^a}=c_0$.
An external electric field $E$ parallel to the walls forces the electroosmotic flow in the channel.
The intensity of the electric field has been tuned to control the electroosmotic velocity.
Finally, the wall charge $\sigma$ is tuned to control the $\zeta$-potential.
After equilibration, a Debye layer sets in at the walls.
Ten systems were simulated five each for the symmetric and antisymmetric setting, corresponding to different values of Debye length and surface charge.
The cationic and anionic density profiles are plotted in Fig.~\ref{fig:charge} together with the particle electrostatic potential for several symmetric and antisymmetric systems in comparison with the predictions of the linearized Poisson-Boltzmann model.
The simulated ionic density is in good agreement with the analytical predictions for all simulations, except for  slight differences for the largest $\zeta$-potentials (see Fig.~\ref{fig:charge}c).
This is not unexpected, since the Debye approximation is bound to fail at large $\beta q\zeta$.

\label{sec:electroosmosis}

\begin{figure}
\centering\includegraphics[width=0.8\linewidth]{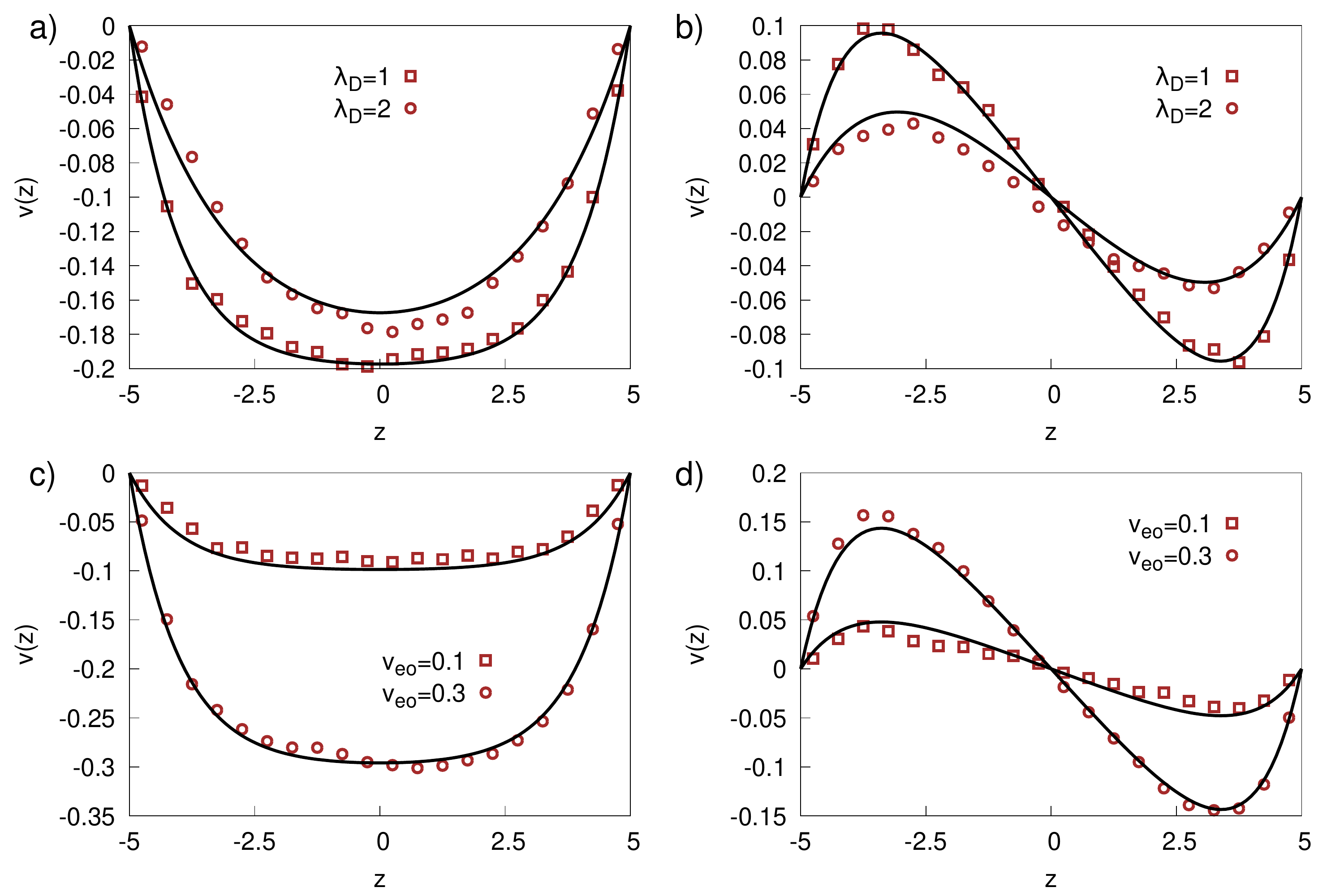}
\caption{
	Velocity profile of the electroosmotic flow generated by a constant electric field parallel to the walls compared with the analytical solutions (black lines). Figures a) and c) correspond to cases with both walls positively charged, while figures b) and d) are cases with a positive charge in the lower wall, $z=-5$, and a negative charge in the upper wall, $z=5$.
In the two top plots, two different values of the Debye length are considered, while in the bottom plots the electroosmotic velocity is changed by changing the electric field and $\lambda_D=1$.
In all the plots, $\zeta=13.7$.
}
\label{fig:velocity}
\end{figure}

The electroosmotic flow generated by the external electric field through the charge imbalance near the walls is plotted in Fig.~\ref{fig:velocity} for different Debye lengths and electroosmotic velocities in comparison with the analytical predictions.

\subsection{Mapping to dimensional units}
\label{sec:parameters}

After the previous discussion on the general features of the proposed approach, 
in this section we consider the specific case of a channel with height
$h = 10 \, {\rm nm}$, Debye length $\lambda_D = 4.28 \, {\rm nm}$, mass density $10^3 \, {\rm Kg/m^3}$, electric conductivity $\kappa = 70 \, {\rm mS/m}$ and viscosity $\mu = 6.5\cdot 10^{-4}\,{\rm Pa\cdot s}$.
The reference dimensional quantities are: length $L_{ref}=0.5\,{\rm nm}$, 
energy $E_{ref}=1.38\cdot10^{-23}\,{\rm J}$, and mass $M_{ref}=3.33\cdot 10^{-25}\,{\rm Kg}$ which corresponds to the mass of a single EH-DPD particle.
The reference charge is set to $Q_{ref}=7.6\cdot10^{-21}\,{\rm C}$, leading to a relative dielectric constant $\epsilon_r = 75$.
The assigned Debye length is obtained by using the (dimensionless) concentration of ions (dimensionless charge $q = 0.3$) $c_0 = 1.875$. 
The particle interaction cutoff is set to $r_c=1\, {\rm nm}$. 
The dimensionless meso-particle density $\rho=0.375$ provides the target mass density of the solution.
 
The remaining physical parameters to be mapped are the solution viscosity and conductivity. This requires preliminary calibration simulations to determine their 
dependence on the model parameters $\gamma$, $\gamma^c_0$ and  $\gamma^a_0$.
In principle, $\gamma^c_0$ and  $\gamma^a_0$ can be used to independently 
reproduce anion and cation conductivities.
Limiting, for simplicity, the analysis to symmetric solutions,  we assume $\gamma^c_0 = \gamma^a_0$.

\begin{figure}
\centering\includegraphics[width=0.8\linewidth]{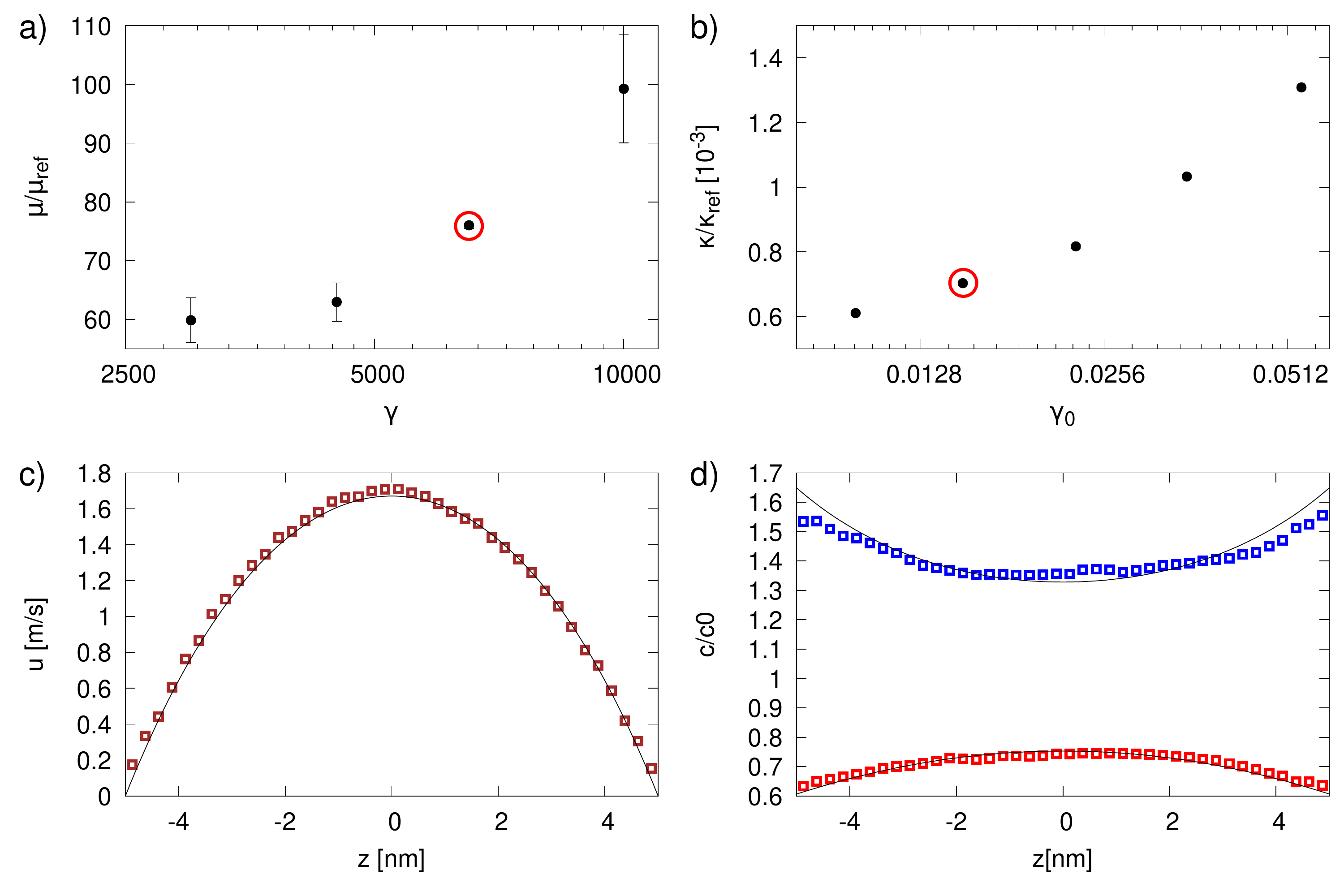}
	\caption{Electroosmotic flow for a $10, {\rm nm}$ planar channel.
	a) Viscosity of the fluid as a function of the parameter $\gamma$, with 
         $\mu_{ref} = M_{ref}^{1/2}E_{ref}^{1/2}L_{ref}^{-2}=8.57\cdot 10^{-6} {\rm Pa\cdot s}$. 
        The value highlighted with a circle was used for the electroosmotic flow simulation.
	b) Conductivity of the fluid as a function of the parameters $\gamma^a=\gamma^c=\gamma_0$, with $\kappa_{ref}=Q_{ref}^2L_{ref}^{-2}E_{ref}^{-1/2}M_{ref}^{-1/2}=107.8\, {\rm S/m}$. The value highlighted with a circle was used for the electroosmotic flow simulation. 	
	c) Velocity and d) density profiles of cations (blue) and anions (red). 
       The Debye length was set to $4.28\, {\rm nm}$.
}
\label{fig:dimensional}
\end{figure}

Figure~\ref{fig:dimensional}a provides the solution viscosity as a function of $\gamma$ for fixed $\gamma^c_0 = \gamma^a_0$. 
We computed the viscosity by imposing a Poiseuille flow~\cite{boromand2015viscosity} and measuring the average velocity obtained at a given pressure difference. 
In the investigated range of 
parameters, the viscosity is found to be almost independent of  $\gamma^c_0$ and $\gamma^a_0$, which control the conductivity.
From the $\mu-\gamma$ curve, we find that $\gamma=6500$ provides the target viscosity.
Analogously, Fig.~\ref{fig:dimensional}b provides the solution conductivity as a function of the common value of $\gamma^c_0 = \gamma^a_0 = \gamma_0$.
From the figure,  $\gamma_0=0.015$ yields the assigned electrical conductivity.
We estimated the electric conductivity by applying a constant electric field to a bulk system and measuring the resulting current.
Moreover, since the model parameters are changed with respect to the  
case reported in Fig~\ref{fig:charge} and \ref{fig:velocity}, the wall model needs to be recalibrated
in order to achieve a no slip boundary condition. Here, we used 
$V_{wall}=6.4\cdot 10^{-4}$ for the inner layer and $V_{wall} = 0.064$ for the external one.
The resulting electroosmotic flow in a $10\, {\rm nm}$ planar channel is shown in Fig.~\ref{fig:dimensional}c, while the ionic density is shown in Fig.~\ref{fig:dimensional}d. The zeta potential is such that $\zeta=-0.5 (q\beta)^{-1}$.
It is apparent that in the middle of the channel, the concentration of anions and cation is different,
as expected when the electric double layers overlap.

\subsection{Excluded volume effects}

\begin{figure}
\centering\includegraphics[width=\linewidth]{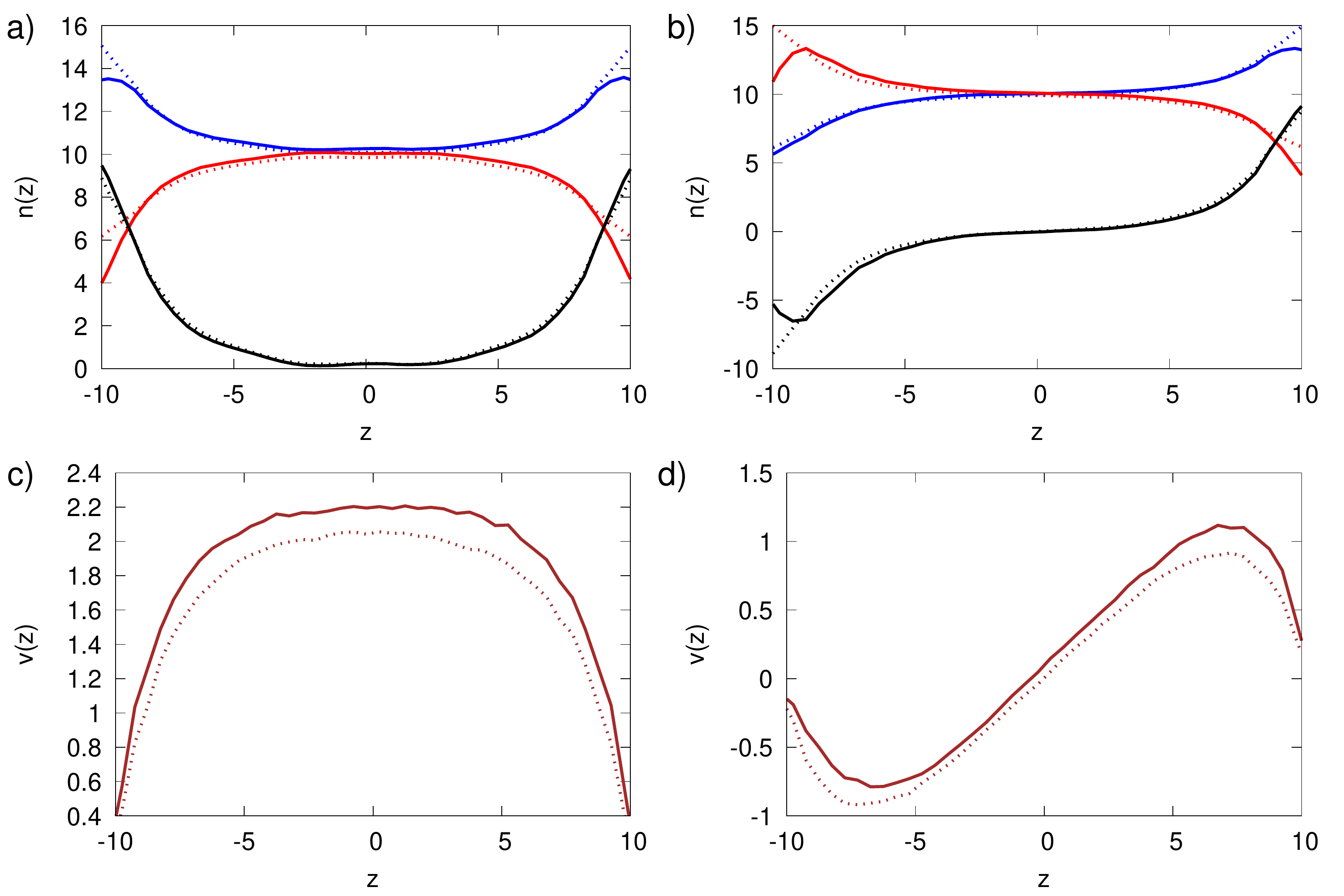}
	\caption{
		Simulation results for a fluid with perfect gas and Van der Waals equation of state. In all the panels, the dashed lines refer to the perfect gas model, while the continuous lines refer to the Van der Waals model of Eq.~\eqref{eq:free_energy-vdw}, with $a=10$, $b^s=0.01$, $b^c=0.02$ and $b^a=0.04$. All the remaining parameters are set as described in Section~\ref{sec:parameters}, with the only difference of $\gamma=4500$ in the perfect gas case to match the viscosity of the corresponding Van der Waals model which has $\gamma=6500$.
       The Debye length was set to $2.15$ in all the cases by tuning the charge of the ions.
	a-b) Cation (blue) and anion (red) concentration profiles and concentration difference (black) for the perfect gas case (continuous lines) and the Van der Waals model (dashed lines), for a planar channel with symmetric 
(a) and antisymmetric (b) charges in the walls.
c-d) Velocity profiles corresponding panel a,
symmetric case (c), and b, antisymmetric case (d).
}
\label{fig:vdw}
\end{figure}

To test the capability of EH-DPD to simulate fluids with different equations of state, we performed planar electroosmosis simulations similar to the above described using a Van der Waals equation of state, \textit{i.e.}
\begin{equation}
	P=\frac{k_BT}{v-b}-\frac{a}{v^2}\;,
\end{equation}
where $v$ is the volume divided by the total number of particles, $a$ is a parameter modeling the attraction between fluid particles and $b$ is the excluded volume of the single particle.
For a multicomponent fluid, the parameters $a$ and $b$ are usually expressed as combinations of the respective single components parameters according to mixing rules~\cite{kwak1986van}. 
In the case of a fluid with three components (\textit{i.e.} $n_c$ cations, $n_a$ anions and $n_s$ solvent), for the meso-particle $i$, $b_i=n^c_i b^c+n^c_a b^a+ n^s_i b^s$ and $a_i=n^c_in^c_i a^{cc}+n^a_in^a_i a^{aa}+2n^s_in^s_i a^{ss}+2n^c_in^a_i a^{ca}+2n^c_in^s_i a^{cs}+2n^s_in^a_i a^{sa}$, where $b^c$, $b^a$ and $b^s$ are the excluded volume of each species and $a^{aa}$, $a^{cc}$, $a^{ss}$, $a^{ca}$, $a^{cs}$ and $a^{sa}$ are parameters controlling the attractive force between the species. 
As in the case of perfect gas, we assumed $n^c+n^a+n^s=M$ to hold for each particle, leaving only the $n^c$ and $n^a$ as independent variables. The free energy is hence
\begin{equation}\label{eq:free_energy-vdw}
	A_i(V_i,n^c_i,n^a_i)=\frac{n^c_i}{\beta}\left[\log\left(\frac{n^c_i}{M-n^c_i-n^a_i}\right)-1\right]+\frac{n^a_i}{\beta}\left[\log\left(\frac{n^a_i}{M-n^c_i-n^a_i}\right)-1\right]+\frac{M}{\beta}\log\left(\frac{M-n^a_i-n^c_i}{V_i-b_i}\right)+\frac{a_i}{V_i^2}\;,
\end{equation}
We chosed to focus on the effect of the corresponding excluded volume parameters $b^s$, $b^c$ and $b^a$, letting the parameter $a=10$ constant and independent on the local fluid composition, such that the fluid is single phase.
In this framework, the chemical potential for cations in particle $i$ reads
\begin{equation}
	\mu^c_i=\frac{1}{\beta}\left[\log\left(\frac{n^c_i}{M-n^c_i-n^a_i}\right)+\frac{M b^c}{V_i-b_i}\right]+q_c\Phi_i\;,
\end{equation}
and similarly for the anions.

The results of the simulations are displayed in Fig.~\ref{fig:vdw}. In Fig~\ref{fig:vdw}a and b, the ionic concentration profiles are shown for two cases each corresponding to the symmetric (panel a) and antisymmetric (panel b) setting.
As reference cases, the perfect gas model above reported is considered (dashed lines), with the same parameters reported in~Section\ref{sec:parameters} with the only difference that $\gamma=4500$, which was adopted to obtain a viscosity equal to the simulated Van der Waals fluid. The applied electric field was such that $v_{eo}=2$.
The continuous lines refer to a Van der Waals model with $b^s=0.01$, $b^c=0.02$ and $b^a=0.04$.
The effect of the excluded volume is to decrease the ion concentration near the walls, generating a larger charge density near negatively charged surfaces and a smaller charge density near positively charged surfaces, due to the larger excluded volume of anions with respect to cations. 
The difference in charge distribution (in particular close to the walls)
 leads to a more pronounced electroosmotic flow in the symmetric case reported in Fig.~\ref{fig:vdw}c, while a net flow is observed in the case of antisymmetrically charged walls due to the different ion accumulation in the two walls, as reported in Fig.~\ref{fig:vdw}d.

\section{Conclusions}

The  EH-DPD (ElectroHydrodynamic Dissipative Particle Dynamics) model we have illustrated in the present paper
is an extension of the DPD model which can be used to simulate the dynamics of  electrolyte solutions at mesoscopic scales. 
The meso-particles carry and exchange among them two ionic species under the collective motion induced by their mutual interactions.
The forces acting between the meso-particles and the ionic exchange rates are determined by the specific fluid model
which, in the EH-DPD, amount to define the free-energy of the meso-particle system.
The model has been validated simulating the electroosmotic flow in a planar nanochannel with charged walls, and a good agreement is obtained both in the case of Debye length comparable with channel height and in the case of small Debye length.
This approach can be used to study fluid systems where thermal fluctuations are crucial on scales larger than affordable with 
 Molecular Dynamics like nanoparticle and biomolecule sensing, systems with membranes for desalination or energy harvesting.
\newtr{We validated our methods against analytical results 
for electroosmotic flows in a planar channel obtained 
from the commonly adopted linearized Poisson-Boltzmann solution of
the Poisson-Nernst-Plank-Navier-Stokes equation.
This was accomplished  using the free energy of a perfect 
gas to model the interparticle interactions.
It has also been shown that different free energies representing more complex fluids can be used. As an example 
we employed the Van der Waals equation of state, which introduces ion specific effects such as the excluded volume.
As a final remark, although in this work we focused on a single phase fluid, the same equation of state allows phase transitions and introduces an energetic cost for the creation of interfaces between different phases allowing to deal with multiphase systems.
In general, the possibility of providing the equation of state as an input of the simulation is promising for the study of systems with ion specific effects driven by electric field or concentration gradients also in presence of phase transitions like, \textit{e.g.}, for hydrophobic channels and hydrophobic nanoporous materials.}

\appendix

\section{Ionic current density}
In this Appendix, we derive an expression for the electric current density of EH-DPD systems.
\label{sec:current}
Consider a system of $N$ EH-DPD particles moving within a volume $V$.
The total charge density of the system is given by
\begin{equation}
\rho(\boldsymbol{r})=\sum\limits_{i=1}^Nq_ig(\boldsymbol{r}-\boldsymbol{x}_i)\;,
\end{equation}
where $g$ is a Gaussian function centered at $\boldsymbol{x}_i$ with constant variance $s$
\begin{equation}
g(\boldsymbol{y})=\frac{1}{\left(2\pi s^2\right)^{3/2}}\exp\left(-\frac{\boldsymbol{y}\cdot\boldsymbol{y}}{2s^2}\right)\;.
\end{equation}
The change of the charge density has two components
\begin{equation}\label{eq:drho}
\mathrm{d}{\rho}(\boldsymbol{r})=\sum\limits_{i=1}^N\left[\mathrm{d}{q}_ig(\boldsymbol{r}_i)+q_i\mathrm{d}{g}(\boldsymbol{r}_i)\right]=\mathrm{d}\rho_1(\boldsymbol{r})+\mathrm{d}\rho_2(\boldsymbol{r})\;,
\end{equation}
where $\boldsymbol{r}_i=\boldsymbol{r}-\boldsymbol{x}_i$.
The first component is
\begin{equation}
\mathrm{d}\rho_1(\boldsymbol{r})=\sum\limits_{i=1}^N\mathrm{d}q_ig(\boldsymbol{r}_i)=\sum\limits_{i=1}^N\left(q_c\mathrm{d}n^c_i-q_a\mathrm{d}n^a_i\right)g(\boldsymbol{r}_i)\;.
\end{equation}
Using Eqs.~\eqref{eq:ehdpd_3} and~\eqref{eq:ehdpd_4}
\begin{equation}
\left\langle\mathrm{d}\rho_1(\boldsymbol{r})\right\rangle=
\sum\limits_{i=1}^N\sum\limits_{j\ne i}\left(q_c\gamma^c_{ij}w^D_{ij} h^c_{ij}-q_a\gamma^a_{ij}w^D_{ij} h^a_{ij}\right)g(\boldsymbol{r}_i)\mathrm{d}t\;.
\end{equation}
Using the antisymmetry of $h^{c/a}_{ij}$
\begin{equation}
\left\langle\mathrm{d}\rho_1(\boldsymbol{r})\right\rangle=
\frac{1}{2}\sum\limits_{i=1}^N\sum\limits_{j\ne i}\left(q_c\gamma^c_{ij}w^D_{ij} h^c_{ij}-q_a\gamma^a_{ij}w^D_{ij} h^a_{ij}\right)\left[g(\boldsymbol{r}_i)-g(\boldsymbol{r}_j)\right]\mathrm{d}t\;.
\end{equation}
We can expand $g(\boldsymbol{r}_j)$ around $\boldsymbol{r}_i$ obtaining
\begin{equation}\label{eq:drho1}
\left\langle\mathrm{d}\rho_1(\boldsymbol{r})\right\rangle=
-\frac{1}{2}\sum\limits_{i=1}^N\sum\limits_{j\ne i}\left(q_c\gamma^c_{ij}w^D_{ij} h^c_{ij}-q_a\gamma^a_{ij}w^D_{ij} h^a_{ij}\right)\left(\sum\limits_{\alpha=1}^\infty\frac{1}{\alpha!}\boldsymbol{\nabla}^\alpha g(\boldsymbol{r}_i):\boldsymbol{x}_{ij}^\alpha\right)\mathrm{d}t\;,
\end{equation}
where $\boldsymbol{\nabla}^\alpha=
\underset{\alpha}{\underbrace{\boldsymbol{\nabla}\otimes\dots\otimes\boldsymbol{\nabla}}}
$, $\boldsymbol{x}_{ij}=\boldsymbol{x}_i-\boldsymbol{x}_j$,
$\boldsymbol{x}_{ij}^\alpha= \underset{\alpha}{\underbrace{\boldsymbol{x}_{ij}\otimes\dots\otimes\boldsymbol{x}_{ij}}}$.
Equation~\eqref{eq:drho1} can be rewritten as
\begin{equation}\label{eq:drho1_bis}
\left\langle\dot{\rho}_1(\boldsymbol{r})\right\rangle=
-\boldsymbol{\nabla}\cdot\left[\frac{1}{2}\sum\limits_{i=1}^N\sum\limits_{j\ne i}\left(q_c\gamma^c_{ij}w^D_{ij} h^c_{ij}-q_a\gamma^a_{ij}w^D_{ij} h^a_{ij}\right)\Gamma_{ij}(\boldsymbol{r})\boldsymbol{x}_{ij}\right]\;,
\end{equation}
where
\begin{equation}
\Gamma_{ij}(\boldsymbol{r})=\left(g(\boldsymbol{r}_i)+\sum\limits_{\alpha=1}^\infty\frac{1}{(\alpha+1)!}\boldsymbol{\nabla}^\alpha g(\boldsymbol{r}_i):\boldsymbol{x}_{ij}^\alpha\right) \ .
\end{equation}
For what concerns the second term in Eq.~\eqref{eq:drho} we have
\begin{eqnarray}
\left\langle\mathrm{d}\rho_2(\boldsymbol{r})\right\rangle=
\sum\limits_{i=1}^N\left(q_cn^c_i-q_an^a_i\right)\mathrm{d}{g}(\boldsymbol{r}_i)=
-\sum\limits_{i=1}^N\left(q_cn^c_i-q_an^a_i\right)\boldsymbol{\nabla}g(\boldsymbol{r}_i)\cdot\boldsymbol{v}_i\mathrm{d}t =
\nonumber \\
- \boldsymbol{\nabla} \cdot  \sum\limits_{i=1}^N\left(q_cn^c_i-q_an^a_i\right)g(\boldsymbol{r}_i)\boldsymbol{v}_i\mathrm{d}t
\;.
\end{eqnarray}
The expected value of the total charge density variation can be written in the form
\begin{equation}
\left\langle\dot{\rho}(\boldsymbol{r})\right\rangle=
-\boldsymbol{\nabla}\cdot\boldsymbol{J}(\boldsymbol{r})\;,
\end{equation}
where $\boldsymbol{J}(\boldsymbol{r})$ is the current density, which reads
\begin{equation}
\boldsymbol{J}(\boldsymbol{r})=\sum\limits_{i=1}^N\left[\frac{1}{2}\sum\limits_{j\ne i}\left(q_c\gamma^c_{ij}w^D_{ij} h^c_{ij}-q_a\gamma^a_{ij}w^D_{ij} h^a_{ij}\right)\Gamma_{ij}(\boldsymbol{r})\boldsymbol{x}_{ij}+\left(q_cn^c_i-q_an^a_i\right)g(\boldsymbol{r}_i)\boldsymbol{v}_i\right]\;.
\end{equation}
In order to obtain the current density $\boldsymbol{J}_i$ associated with particle $i$ (hereafter called the particle current density), $\boldsymbol{J}$ is integrated over the whole space, i.e.
\begin{equation}
\sum\limits_{i=1}^N\boldsymbol{J}_iV_i=\int\boldsymbol{J}\;\mathrm{d}\boldsymbol{r}=\sum\limits_{i=1}^N\left[\frac{1}{2}\sum\limits_{j\ne i}\left(q_c\gamma^c_{ij}w^D_{ij} h^c_{ij}-q_a\gamma^a_{ij}w^D_{ij} h^a_{ij}\right)\boldsymbol{x}_{ij}+\left(q_cn^c_i-q_an^a_i\right)\boldsymbol{v}_i\right]\;,
\end{equation}
which defines the particle current density as
\begin{equation}
\boldsymbol{J}_i=\frac{1}{2V_i}\sum\limits_{j\ne i}\left(q_c\gamma^c_{ij}w^D_{ij} h^c_{ij}-q_a\gamma^a_{ij}w^D_{ij} h^a_{ij}\right)\boldsymbol{x}_{ij}+\frac{1}{V_i}\left(q_cn^c_i-q_an^a_i\right)\boldsymbol{v}_i\;.
\end{equation}
The average current density given the state of the system is hence
\begin{equation}\label{eq:current_density}
\boldsymbol{J}(\boldsymbol{x},\boldsymbol{v},\boldsymbol{n^c}\boldsymbol{n^a})=\frac{1}{V}\sum\limits_{i=1}^N\boldsymbol{J}_iV_i=\frac{1}{V}\sum\limits_{i=1}^N\left[\frac{1}{2}\sum\limits_{j\ne i}\left(q_c\gamma^c_{ij}w^D_{ij} h^c_{ij}-q_a\gamma^a_{ij}w^D_{ij} h^a_{ij}\right)\boldsymbol{x}_{ij}+\left(q_cn^c_i-q_an^a_i\right)\boldsymbol{v}_i\right] \ .
\end{equation}
We used this expression to compute the current density of several bulk EH-DPD systems when an external electric field is applied, in order to evaluate the conductivity of the solution.

\section{Electrical conductivity of the system}
\label{sec:conductivity}
Here, we  derive the expression for the conductivity of an EH-DPD system in the context of a linearized theory \textit{\`a la} Green-Kubo~\cite{marconi2008fluctuation}.
The time dependent pdf of the state of a system of EH-DPD particles at equilibrium in the infinite past
will obey the initial condition
\begin{equation}
P(\boldsymbol{y},-\infty)=P_{eq}(\boldsymbol{y})=\frac{1}{Z}\exp\left(\frac{S_{eq}(\boldsymbol{y})}{k_B}\right)\;,
\end{equation} 
where $\boldsymbol{y}$ is the state of the system and $Z$ the normalization constant. 
At time $t=-\infty$ an external electric field $\boldsymbol{E}(t)=E(t)\boldsymbol{\hat{n}}$ is switched on.
The pdf along the successive evolution driven by the external electric field is governed by the Fokker-Planck equation, Eq.~\eqref{eq:fokkerplanck}, repeated here for convenience,
\begin{equation}\label{eq:fokkerplanck2}
\frac{\partial P(\boldsymbol{y},t)}{\partial t}=
-\boldsymbol{\nabla}\cdot\left[\left(\boldsymbol{u}^C(\boldsymbol{y})+\frac{1}{k_B}\boldsymbol{D}(\boldsymbol{y})\cdot\boldsymbol{\nabla}S(\boldsymbol{y})-\boldsymbol{D}(\boldsymbol{y})\cdot\boldsymbol{\nabla}\right)P(\boldsymbol{y},t)\right]=
L_{FP}P(\boldsymbol{y},t)\;,
\end{equation}
where now the Fokker-Planck operator $L_{FP}$ includes the effects of the electric field. 
Specifically, the 
entropy $S(\boldsymbol{y})$, Eq.~\eqref{eq:entropy} accounts for the external perturbation $U_{ext}=-\sum q_i\boldsymbol{x}_i\cdot\boldsymbol{\hat{n}}E(t)$ added to
the electrostatic energy, Eq.~\eqref{eq:electrostatic_energy}.
The total entropy of the system can be expressed as
\begin{equation}\label{eq:s_ne}
S(\boldsymbol{y})=S_{eq}(\boldsymbol{y})+\frac{E(t)\boldsymbol{\hat{n}}}{T}\cdot\sum\limits_{i=1}^Nq_i\boldsymbol{x}_i=S_{eq}(\boldsymbol{y})+S_{ext}(\boldsymbol{y})\;,
\end{equation}
where the external term is proportional to the electric field intensity E(t).
The electric field also enters in the conservative drift $\boldsymbol{u^C}$, Eq.~\eqref{eq:conservative_drift} through the entropy,  Eq.~\eqref{eq:cond_S_1}, and splits into its equilibrium and external components
\begin{equation}\label{eq:u_ne}
\boldsymbol{u^C}(\boldsymbol{y})=\boldsymbol{u^C_{eq}}(\boldsymbol{y})+
\begin{pmatrix}\boldsymbol{0}\\q_1\boldsymbol{\hat{n}}\\\dots\\ q_N\boldsymbol{\hat{n}}\\\boldsymbol{0}\\\boldsymbol{0}
\end{pmatrix}E(t)=\boldsymbol{u^C_{eq}}(\boldsymbol{y})+\boldsymbol{u^C_{ext}}(\boldsymbol{y}) \; .
\end{equation}
Hence, the operator $L_{FP}$ decomposes into an equilibrium component, independent of the electric field $E$, and an external, field-dependent part
\begin{equation}
L_{FP}=L^{eq}_{FP}+L^{ext}_{FP}\;,
\end{equation}
where
\begin{equation}\label{eq:psi_eq}
L^{eq}_{FP}P=-\boldsymbol{\nabla}\cdot\left[\left(\boldsymbol{u^C_{eq}}(\boldsymbol{y})+\frac{1}{k_B}\boldsymbol{D}(\boldsymbol{y})\cdot\boldsymbol{\nabla}S_{eq}(\boldsymbol{y})-\boldsymbol{D}(\boldsymbol{y})\cdot\boldsymbol{\nabla}\right)P\right]\;,
\end{equation}
and
\begin{equation}\label{eq:psi_ne}
L_{FP}^{ext}P=-\boldsymbol{\nabla}\cdot\left[\left(\boldsymbol{u^C_{ext}}(\boldsymbol{y})+\frac{1}{k_B}\boldsymbol{D}(\boldsymbol{y})\cdot\boldsymbol{\nabla}S_{ext}(\boldsymbol{y})\right)P\right]\;.
\end{equation}
Linearizing $P(\boldsymbol{y},t;E)$ for small external fields,
\begin{equation}\label{eq:psplit}
P(\boldsymbol{y},t)=P_{eq}(\boldsymbol{y})+\frac{\mathrm{d} P(\boldsymbol{y},t)}{\mathrm{d} E}E+o(E)\simeq P_{eq}(\boldsymbol{y})+P_{ext}(\boldsymbol{y},t)\;,
\end{equation}
the solution of the Fokker-Planck equation,
\begin{equation}
\frac{\partial P(\boldsymbol{y},t)}{\partial t}=\frac{\partial P_{ext}(\boldsymbol{y},t)}{\partial t}=L_{FP}^{eq}P_{ext}(\boldsymbol{y},t)+L_{FP}^{ext}(t)P_{eq}(\boldsymbol{y})\;,
\end{equation}
where we used $L_{FP}^{eq}P_{eq}=0$, is
\begin{equation}\label{eq:p_ne_solution1}
P_{ext}(\boldsymbol{y},t)=\int_{-\infty}^t\exp\left[\left(t-s\right)L_{FP}^{eq}\right]L_{FP}^{ext}(s)P_{eq}\,\mathrm{d}s\;.
\end{equation}
The action of the operator $L_{FP}^{ext}$ on the equilibrium pdf can be expressed as (Eq.~\eqref{eq:psi_ne})
\begin{equation}
L_{FP}^{ext}P_{eq}(\boldsymbol{y})=-\frac{1}{k_B}
\left[\boldsymbol{u^C_{ext}}(\boldsymbol{y})\cdot\boldsymbol{\nabla}S_{eq}(\boldsymbol{y})+\boldsymbol{\nabla}\cdot\boldsymbol{D}(\boldsymbol{y})\cdot\boldsymbol{\nabla}S_{ext}(\boldsymbol{y})+\frac{1}{k_B}\boldsymbol{\nabla}S_{ext}(\boldsymbol{y})\cdot\boldsymbol{D}(\boldsymbol{y})\cdot\boldsymbol{\nabla}S_{eq}(\boldsymbol{y})\right]P_{eq}(\boldsymbol{y})\;,
\end{equation}
where  $\boldsymbol{\nabla}\cdot\boldsymbol{u^C_{ext}}=0$ and $\boldsymbol{D}:\boldsymbol{\nabla}\otimes\boldsymbol{\nabla}S_{ext}=0$.
According to the definitions of $\boldsymbol{D}$ (Eqs.~\ref{eq:d_blocks_v}-\ref{eq:d_blocks_a}), $\boldsymbol{u}^C_{ext}$ (Eq.~\ref{eq:u_ne}), $S_{eq}$ and $S_{ext}$ (Eq.~\ref{eq:u_ne}), the action of $L^{ext}_{FP}$ on $P_{eq}$ reduces to
\begin{equation}\label{eq:phi2}
L_{FP}^{ext}P_{eq}(\boldsymbol{y})=
\beta\sum\limits_{i=1}^N\left[-\frac{1}{2}\sum\limits_{j\ne 1}^N\left(q^c\gamma^c_{ij}w^D_{ij}h^c_{ij}-q^a\gamma^a_{ij}w^D_{ij}h^a_{ij}\right)\boldsymbol{x}_{ij}+q_i\boldsymbol{v}_i\right]\cdot\boldsymbol{\hat{n}} E(t) P_{eq}(\boldsymbol{y})=-\beta V \boldsymbol{J}^-_{eq}\cdot\boldsymbol{\hat{n}}E(s)P_{eq}(\boldsymbol{y})  \; ,
\end{equation}
where the function $\boldsymbol{J}^-_{eq}(\boldsymbol{y})$ is obtained computing the current density of Eq.~\eqref{eq:current_density} in absence of any external field, changing $\boldsymbol{v}$ in $-\boldsymbol{v}$. 
We obtain
\begin{equation}\label{eq:p_ne_solution2}
P_{ext}(\boldsymbol{y},t)=-\beta V  \int\limits_{-\infty}^t
\exp\left[(t-s)L_{FP}^{eq}\right]\left[\boldsymbol{J}_{eq}^-\cdot\boldsymbol{\hat{n}}E(s) P_{eq}(\boldsymbol{y})\right]
\mathrm{d}s\; .
\end{equation}
With this expression, the current density due to a stationary electric field after the transient decayed, e.g., at time $0$ is
\begin{equation}\label{eq:p_ne_solution3}
\left\langle\boldsymbol{J}\right\rangle=
\int \boldsymbol{J} P(\boldsymbol{y},0)\mathrm{d}\boldsymbol{y}=
\int \boldsymbol{J} P_{eq}(\boldsymbol{y})\mathrm{d}\boldsymbol{y}+
\int \boldsymbol{J} P_{ext}(\boldsymbol{y},0)\mathrm{d}\boldsymbol{y}=
\boldsymbol{J_1}+\boldsymbol{J_2}\;.
\end{equation}
$\boldsymbol{J_1}$ can be simplified by considering that the current density
\begin{equation}
\boldsymbol{J}(\boldsymbol{y})=
\frac{1}{V}\sum\limits_i\left[
\frac{1}{2}\sum\limits_{j\ne 1}^N\left(q^c\gamma^c_{ij}w^D_{ij}h^c_{ij}-q^a\gamma^a_{ij}w^D_{ij}h^a_{ij}\right)\boldsymbol{x}_{ij}+q_i\boldsymbol{v}_i+E\frac{1}{2}\sum\limits_{j\ne 1}^N\left(q_c^2\gamma^c_{ij}w^D_{ij}
+q_a^2\gamma^a_{ij}w^D_{ij}\right)\boldsymbol{x}_{ij}\otimes\boldsymbol{x}_{ij}\cdot\boldsymbol{\hat{n}}\right]
\end{equation} 
can be decomposed into equilibrium and external contributions
\begin{equation}
\boldsymbol{J}(\boldsymbol{y})=
\boldsymbol{J}_{eq}(\boldsymbol{y})+\boldsymbol{J}_{ext}(\boldsymbol{y})\;.
\end{equation} 
Considering that
\begin{equation}
\int\boldsymbol{J}_{eq}(\boldsymbol{y})P_{eq}(\boldsymbol{y})\mathrm{d}\boldsymbol{y}=0\;,
\end{equation}
$\boldsymbol{J_1}$ reads
\begin{equation}
\boldsymbol{J_1}=
\int\boldsymbol{J}_{ext}(\boldsymbol{y})P_{eq}(\boldsymbol{y})\mathrm{d}\boldsymbol{y}=
\frac{E}{2V}\sum\limits_i
\sum\limits_{j\ne 1}^N\left\langle\left(q_c^2\gamma^c_{ij}w^D_{ij}+q_a^2\gamma^a_{ij}w^D_{ij}\right)\boldsymbol{x}_{ij}\otimes\boldsymbol{x}_{ij}\right\rangle_{eq}\cdot\boldsymbol{\hat{n}}\;.
\end{equation}
The second component $\boldsymbol{J_2}$ is
\begin{equation}
\boldsymbol{J_2}=
\int\boldsymbol{J}(\boldsymbol{y})P_{ext}(\boldsymbol{y},0)\mathrm{d}\boldsymbol{y}=
\int\left(\boldsymbol{J}_{eq}(\boldsymbol{y})+\boldsymbol{J}_{ext}(\boldsymbol{y})\right)P_{ext}(\boldsymbol{y},0)\mathrm{d}\boldsymbol{y}\; .
\end{equation}
Considering that $\boldsymbol{J}_{ext}P_{ext}=O(E^2)$, the linearized expression for $\boldsymbol{J_2}$ is
\begin{equation}
\boldsymbol{J_2}=
\int\boldsymbol{J}_{eq}(\boldsymbol{y})P_{ext}(\boldsymbol{y},0)\mathrm{d}\boldsymbol{y}=-\beta VE\int\limits_{-\infty}^0\int
\boldsymbol{J}_{eq}(\boldsymbol{y})\otimes\exp\left[-sL_{FP}^{eq}\right]\left[\boldsymbol{J}_{eq}^-\cdot \boldsymbol{\hat{n}} P_{eq}(\boldsymbol{y})\right]
\mathrm{d}s\mathrm{d}\boldsymbol{y}\; .
\end{equation}
Using the adjoint operator $\exp\left(-s L_{FP}^{eq^\dag}\right)$
\begin{equation}
\boldsymbol{J_2}=
\beta VE\int\limits_{-\infty}^0\int\exp\left(-sL_{FP}^{eq^\dag}\right)
\left[\boldsymbol{J}_{eq}(\boldsymbol{y})\right]\otimes\boldsymbol{J}_{eq}^-\cdot\boldsymbol{\hat{n}} P_{eq}(\boldsymbol{y})
\mathrm{d}s\mathrm{d}\boldsymbol{y}\; ,
\end{equation}
which gives
\begin{equation}
\boldsymbol{J_2}=
-\beta V\int\limits^{\infty}_0
\left\langle\boldsymbol{J}(s)\otimes \boldsymbol{J}^-(0)\right\rangle_{eq}
\mathrm{d}s\cdot E\boldsymbol{\hat{n}}\;.
\end{equation}
Introducing the conductivity tensor $\boldsymbol{\kappa}$ such that $\boldsymbol{J}=\boldsymbol{\kappa}\cdot\boldsymbol{E}$,
\begin{equation}
\boldsymbol{\kappa}=
\frac{1}{2V}\sum\limits_i 
\sum\limits_{j\ne 1}^N\left\langle\left(q_c^2\gamma^c_{ij}w^D_{ij}+q_a^2\gamma^a_{ij}w^D_{ij}\right)\boldsymbol{x}_{ij}\otimes\boldsymbol{x}_{ij}\right\rangle_{eq}-
\beta V \int\limits_{0}^\infty\left\langle
\boldsymbol{J}(s)\boldsymbol{J}^-(0) \right\rangle_{eq}\mathrm{d}s\;.
\end{equation}
Since the equilibrium system is statistically isotropic it follows $\kappa_{ij}=\kappa_0\delta_{ij}$.
We used the derived formula as an additional tool to evaluate the conductivity of the EH-DPD system, finding a good agreement with the conductivity computed in nonequilibrium simulations with an extrnal electric field applied.

\bibliography{articoli}
\bibliographystyle{unsrt}

\end{document}